\pdfoutput=1
\documentclass[preprint,12pt]{elsarticle}

\usepackage{graphicx}
\usepackage{subcaption}
\usepackage{multirow}
\usepackage{amsmath,amssymb,amsfonts}
\usepackage{xcolor}
\usepackage{textcomp}
\usepackage{booktabs}
\usepackage{rotating}
\usepackage{algorithm}
\usepackage{algorithmicx}
\usepackage{algpseudocode}
\usepackage{float}
\usepackage{placeins}
\usepackage{microtype}
\usepackage[hidelinks]{hyperref}

\graphicspath{{pic/}}

\makeatletter
\def\ps@pprintTitle{%
  \let\@oddhead\@empty
  \let\@evenhead\@empty
  \let\@oddfoot\@empty
  \let\@evenfoot\@empty}
\makeatother

\usepackage[normalem]{ulem}
\newif\ifmarkchanges
\markchangesfalse
\ifmarkchanges
  
  \newcommand{\del}[1]{{\color{red}\sout{#1}}}
\else
  
  \newcommand{\del}[1]{}
\fi

\begin{document}

\begin{frontmatter}

\title{Coupled Thermochemical Ablation and Rarefied Particle Erosion during Mars Entry}

\author[ucf]{Hyeon Woo Nam}
\author[ucf]{Sung Min Jo\corref{cor1}}
\ead{sungmin.jo@ucf.edu}
\cortext[cor1]{Corresponding author}
\affiliation[ucf]{organization={Center of Excellence in Hypersonics and Space Propulsion (HYPERSPACE), Department of Mechanical and Aerospace Engineering, University of Central Florida},
  city={Orlando},
  postcode={32816},
  state={FL},
  country={USA}}

\begin{abstract}
Martian dust particles are a few micrometers in diameter, comparable to the mean free path of the gas in the hypersonic shock layer. Their Knudsen number ranges from approximately 0.2 to 100, placing them in the transitional and free-molecular regimes, whereas the flow around the vehicle is a continuum. The two-way flow-particle coupling previously developed by the authors for a two-temperature thermochemical nonequilibrium flow is extended to a receding and pyrolyzing heat shield surface. The gas-surface interaction and the in-depth material response are solved on separate time scales. The surface balances are exchanged at every flow step, while the in-depth response is coupled implicitly over physical time windows. The flow-material coupling is assessed against graphite arc-jet measurements. The framework is applied to the Schiaparelli trajectory from 50 to 30~km. Three surface chemistry models (equilibrium, Park-5, and Zhluktov--Abe) are compared for fixed geometry, char-induced recession, and combined char ablation and particle erosion.  At the final trajectory point, the stagnation point heat flux differs by 30--75\% among the chemistry models. For the equilibrium model, surface recession changes this heat flux by less than 1\% but lowers wall temperature and increases pyrolysis gas injection. Particle erosion contributes 1.73~mm of stagnation point recession, exceeding the char-induced recession, and more than doubles the total recession for all three models. To the authors' knowledge, this study provides the first quantitative comparison of particle erosion between rarefied and continuum drag correlations, and the continuum correlation underestimates the stagnation point erosion rate by 31--35\%.
\end{abstract}


\end{frontmatter}

\section{Introduction}\label{sec:intro}

Hypersonic atmospheric entry involves coupled fluid dynamics, chemical reactions, energy exchange, radiative transfer, and gas-surface interactions. Strong bow shocks heat the surrounding gas, causing molecular excitation, dissociation, and, at sufficiently high temperatures, ionization   \cite{barbante2004fundamentals,gnoffo1999planetary}. The gas may remain out of thermal and chemical equilibrium, making the predicted heating sensitive to how chemical reactions and internal energy exchange are treated~\cite{barbante2004fundamentals,gnoffo1999planetary,jeong2026first}. Gases released from an ablating surface further modify the boundary layer composition, diffusive energy transport, and radiative emission and absorption \cite{johnston2013influence, jo2026coupled, le2025numerical}. Accurate heating predictions require accounting for interactions between the surrounding flow and the exposed material.

Ablative thermal protection systems (TPS) protect the underlying structure through insulation, thermal decomposition, and surface mass loss. In porous carbon-phenolic materials, resin decomposition generates pyrolysis gases that flow through the material and enter the boundary layer. The resulting response depends on heat conduction, decomposition kinetics, and energy transport within the porous medium \cite{meurisse2018multidimensional,scandelli2023two}. Surface reactions determine wall composition and char ablation rate, influencing both blowing and species diffusion heating \cite{johnston2013influence,zibitsker2023finite}. The material retains its preceding heating and decomposition history, and surface recession changes the exposed geometry and remaining TPS thickness \cite{meurisse2018multidimensional, kuntz2001predictions, martin2015strongly}. These processes are related to surface heating, in-depth temperature, and cumulative material loss.

 During Martian entry, atmospheric dust introduces additional thermal and mechanical effects. As particles traverse the shock layer, they exchange momentum and energy with the gas, while impacts on the heat shield contribute to collision heating and mechanical erosion. The strength of these interactions depends on particle size, material properties, mass loading, and flow conditions \cite{sahai2022variable,hinkle2024efficient}. Martian dust particles are a few micrometers in diameter, which is comparable to the mean free path of the gas in the hypersonic shock layer. The flow around the vehicle is in the continuum regime, whereas the flow around each particle is in the transitional or free-molecular regime. Particle drag and heat transfer are then evaluated with closures that account for rarefaction and compressibility at the particle scale~\cite{henderson1976drag,loth2008compressibility,palmer2023effect}. Under representative Martian dust conditions, mechanical erosion can contribute significantly to TPS recession even when particle-induced changes in surface heating are relatively small \cite{nam2026euler}. Consequently, both heating and erosion must be considered when assessing dust effects on heat shield performance.

Coupled numerical methods have been developed to account for these interactions. Flow-material approaches exchange surface heat flux and wall states through mass and energy balances, with mesh motion included when geometric recession is resolved   \cite{kuntz2001predictions,martin2015strongly, thornton2023coupling}. For particle-laden flows, Euler--Lagrange methods resolve particle trajectories and their exchange of momentum and energy with the carrier gas, with different levels of coupling fidelity \cite{ching2020two,sahai2022variable}. Particle-flux-based trajectory control volume methods provide an efficient means of evaluating surface erosion and volumetric particle source terms~\cite{hinkle2022efficient,hinkle2024efficient}.   These developments provide the basis for examining gas-particle interactions together with transient material response and surface recession. However, to the authors' knowledge, previous studies have treated flow-material and flow-particle coupling separately, which limits their ability to capture the combined effects of ablation and particle erosion on heat shield response.

In this study, a coupled flow-material-particle framework is developed to investigate how surface chemistry and particle erosion affect the thermal response and recession of an ablating heat shield. The particle model of Nam et al.~\cite{nam2026euler}, which couples the particles in both directions with a two-temperature thermochemical nonequilibrium flow, is extended here to a receding and pyrolyzing heat shield surface. The surface balances and the in-depth material response are solved separately on their own time scales, with the surface balances exchanged with the flow at every flow step and the in-depth response iterated implicitly over physical time windows. The particle Knudsen number is evaluated along the trajectory to identify the flow regime at the particle scale and to assess the drag and heat transfer closures in that regime. The framework is demonstrated along the Schiaparelli entry trajectory, with the material heating, decomposition, and recession history carried across consecutive trajectory segments.

The framework combines HEGEL and PLATO   \cite{munafo2024hegel,munafo2025plato} with the OpenFOAM-based ORACLE and PATO solvers \cite{weller1998tensorial,nam2026euler,   meurisse2018multidimensional}   through preCICE~\cite{bungartz2016precice}. Separate PATO participants evaluate the surface balances and the transient in-depth material response. The flow-material coupling is validated against graphite arc-jet measurements and compared with a reference simulation. The framework is then applied to the Schiaparelli trajectory from 50 to 30~km using TACOT as a reference ablative material~\cite{lachaud2014tacot}. Three surface chemistry models (equilibrium~\cite{meurisse2018multidimensional,le2025numerical}, Park-5~\cite{johnston2014study,le2025numerical}, and Zhluktov--Abe~\cite{zhluktov1999viscous}) are combined with three configurations: a fixed surface without particles, a surface receding through char ablation without particles, and a surface receding through both char ablation and particle erosion. 

The objective of this study is to assess how surface chemistry, char-induced recession, and particle erosion affect heat shield performance in a dusty entry environment. Their effects are evaluated by comparing wall heat flux, temperature, mass injection, and cumulative recession. The remainder of this paper is organized as follows. Section~\ref{sec:methods} presents the physical models for each domain, the coupling strategy, and the framework validation. Section~\ref{sec:results} describes the computational conditions and discusses the coupled responses along the selected trajectory.   Section~\ref{sec:conclusion} summarizes the main findings.

\section{Numerical Methods}\label{sec:methods}

\subsection{Nonequilibrium Flow Solver}
  \label{subsec:methods_noneq}
  The gas flow is computed using HEGEL \cite{munafo2024hegel}, an in-house, cell-centered finite-volume solver for chemically and thermally nonequilibrium flows. Thermodynamic and transport properties,
  chemical production rates, and internal energy transfer terms are evaluated using the PLATO library \cite{munafo2025plato}. The two-temperature formulation solves the conservation equations for species mass, mixture momentum, total energy, and vibrational-electronic energy:
  \begin{equation}
      \frac{\partial \rho_i^g}{\partial t}
      +
      \nabla \cdot
      \left(
        \rho_i^g \boldsymbol{u}^g+\boldsymbol{J}_i^g
      \right)
      =
      s_i^g,
      \label{eq:flow_species}
  \end{equation}
  \begin{equation}
      \frac{\partial (\rho^g\boldsymbol{u}^g)}{\partial t}
      +
      \nabla \cdot
      \left(
        \rho^g\boldsymbol{u}^g\otimes\boldsymbol{u}^g
      \right)
      +
      \nabla p^g
      -
      \nabla\cdot\boldsymbol{\tau}^g
      =
      \boldsymbol{\Omega}_{\mathrm{M}}^p,
      \label{eq:flow_momentum}
  \end{equation}
  \begin{equation}
      \begin{split}
      \frac{\partial (\rho^g E^g)}{\partial t}
      &+
      \nabla\cdot
      \left(
        \rho^g H^g\boldsymbol{u}^g
        +\sum_i h_i^g\boldsymbol{J}_i^g
      \right)
      \\
      &=
      \nabla\cdot
      \left(\boldsymbol{\tau}^g\cdot\boldsymbol{u}^g\right)
      -
      \nabla\cdot\boldsymbol{q}^g
      +
      \Omega_{\mathrm{E}}^p,
      \end{split}
      \label{eq:flow_energy}
  \end{equation}
  \begin{equation}
      \begin{split}
      \frac{\partial (\rho^g e_{\mathrm{ve}}^g)}{\partial t}
      &+
      \nabla\cdot
      \left(
        \rho^g e_{\mathrm{ve}}^g\boldsymbol{u}^g
        +\sum_i h_{\mathrm{ve},i}^g\boldsymbol{J}_i^g
      \right)
      \\
      &=
      -p_e^g\nabla\cdot\boldsymbol{u}^g
      -
      \nabla\cdot\boldsymbol{q}_{\mathrm{ve}}^g
      +
      \Omega_{\mathrm{CE}}^g
      +
      \Omega_{\mathrm{VT}}^g.
      \end{split}
      \label{eq:flow_internal_energy}
  \end{equation}
  Here, $\rho_i^g$ and $\rho^g$ are the species and mixture densities, $\boldsymbol{u}^g$ is the mass-averaged velocity, and $\boldsymbol{J}_i^g$ is the species diffusive mass flux. The quantities $p^g$, $\boldsymbol{\tau}^g$, $E^g$, $H^g$, and $\boldsymbol{q}^g$ denote pressure, viscous stress, specific total energy, specific total enthalpy, and conductive heat flux, respectively. The subscript $\mathrm{ve}$ identifies the internal energy mode associated with the second temperature. Because ionization is not considered in this study, the electron pressure work, $-p_e^g\nabla\cdot\boldsymbol{u}^g$, vanishes. The terms $\Omega_{\mathrm{VT}}^g$ and $\Omega_{\mathrm{CE}}^g$ represent the vibrational-translational energy exchange and the vibrational energy lost in heavy-particle dissociation reactions, respectively. The latter is modeled with the nonpreferential dissociation model~\cite{chulpark_textbook}.
  Chemical production and energy transfer terms are evaluated using PLATO, while ORACLE \cite{nam2026euler}, which is a particle solver, supplies the particle momentum and energy sources, $\boldsymbol{\Omega}_{\mathrm{M}}^p$ and $\Omega_{\mathrm{E}}^p$.
  Details of the governing formulation and source terms are provided in Refs.~\cite{munafo2020computational,nam2026euler}. Radiative heat transfer is not included in the present calculations, since the radiative heating of the forebody is negligible compared with the convective heating at the entry velocities of 2.9--5.5~km/s considered here~\cite{johnston2012shock}.

  At the blowing wall, the coupled surface and material calculations provide the temperature $T_w$, species mass fractions $Y_{w,i}$, and blowing mass flux $\dot{m}_w''=\dot{m}_c''+\dot{m}_g''$.
 The blowing boundary condition follows the formulation of Thompson et al.~\cite{thompson2008implementation}. For a prescribed wall temperature, composition, and blowing mass flux, the wall state is determined from the normal momentum balance, mass conservation, and equation of state:
  \begin{equation}
      \rho_w
      =
      \frac{
        p_{\eta}+
        \sqrt{p_{\eta}^{2}-4R_wT_w(\dot{m}_w'')^{2}}
      }{2R_wT_w},
      \label{eq:blowing_density}
  \end{equation}
  \begin{equation}
      u_{n,w}
      =
      \frac{
        2R_wT_w\dot{m}_w''
      }{
        p_{\eta}+
        \sqrt{p_{\eta}^{2}-4R_wT_w(\dot{m}_w'')^{2}}
      },
      \label{eq:blowing_velocity}
  \end{equation}
  \begin{equation}
      p_w
      =
      \frac{
        p_{\eta}+
        \sqrt{p_{\eta}^{2}-4R_wT_w(\dot{m}_w'')^{2}}
      }{2},
      \label{eq:blowing_pressure}
  \end{equation}
  where
  $p_{\eta}=p_{\mathrm{nc}}
  +\rho_{\mathrm{nc}}u_{n,\mathrm{nc}}^{2}$
  is the normal momentum flux evaluated from the adjacent gas cell, denoted by $\mathrm{nc}$. The quantity $R_w$ is the specific gas constant of the wall gas mixture.
  Viscous stresses are neglected in this normal momentum balance. The normal velocity is positive toward the gas. These quantities define the blowing boundary state. HEGEL returns the wall pressure, heat flux, and species diffusive fluxes required by the surface balances.

\subsection{High-Enthalpy Particle Solver}
\label{subsec:methods_particle}

Particle tracking is computed using ORACLE, an OpenFOAM-based Lagrangian solver. The particle formulation follows the Euler--Lagrangian framework developed by Nam et al.~\cite{nam2026euler}. Particles are treated as spherical, with a uniform internal temperature, and interparticle collisions are neglected. Their translational motion is governed by aerodynamic drag:
\begin{equation}
    \frac{d\boldsymbol{x}_p}{dt} = \boldsymbol{v}_p,
    \qquad
    m_p \frac{d\boldsymbol{v}_p}{dt}
    =
    \frac{1}{2}\rho^g C_D A_p
    \left|\boldsymbol{u}^g-\boldsymbol{v}_p\right|
    \left(\boldsymbol{u}^g-\boldsymbol{v}_p\right),
    \label{eq:particle_motion}
\end{equation}
where $m_p$, $\boldsymbol{x}_p$, and $\boldsymbol{v}_p$ are the particle mass, position, and velocity, respectively. The particle cross-sectional area is $A_p=\pi d_p^2/4$, where $d_p$ is the particle diameter. The gas density and velocity are denoted by $\rho^g$ and $\boldsymbol{u}^g$. The drag coefficient $C_D$ is evaluated using the Henderson correlation \cite{henderson1976drag}.

Heat exchange with the thermally nonequilibrium gas accounts for the translational and vibrational-electronic energy modes. The particle temperature is calculated by
\begin{equation}
    m_p c_{p,p}\frac{dT_p}{dt}
    =
    \dot{Q}_p
    =
    \pi d_p \mathrm{Nu}
    \left[
      \kappa_{\mathrm{tr}}^g(T_{\mathrm{tr}}^g-T_p)
      +
      \kappa_{\mathrm{ve}}^g(T_{\mathrm{ve}}^g-T_p)
    \right],
    \label{eq:particle_heating}
\end{equation}
where $c_{p,p}$ is the particle specific heat, and $\kappa_{\mathrm{tr}}^g$ and $\kappa_{\mathrm{ve}}^g$ are the gas thermal conductivities associated with the two energy modes. The Nusselt number $\mathrm{Nu}$ is evaluated using the Fox correlation \cite{fox1978shock}. Particle vaporization is modeled using the vaporization temperature relation of Schaefer et al. \cite{schaefer2004thermodynamic}. During vaporization, heat supplied for phase change produces mass loss, with the corresponding diameter change given by
\begin{equation}
    \frac{dd_p}{dt}
    =
    \frac{2\dot{m}_p}{\pi \rho_p d_p^2},
    \label{eq:particle_diameter}
\end{equation}
where $\rho_p$ is the particle material density and $\dot{m}_p=dm_p/dt$ represents the rate of change in particle mass due to vaporization. The transport of vaporized particle species in the gas phase is not included.

\begin{figure}[htbp]
  \centering
  \begin{subfigure}[t]{0.32\textwidth}
    \centering
    \includegraphics[width=\linewidth]{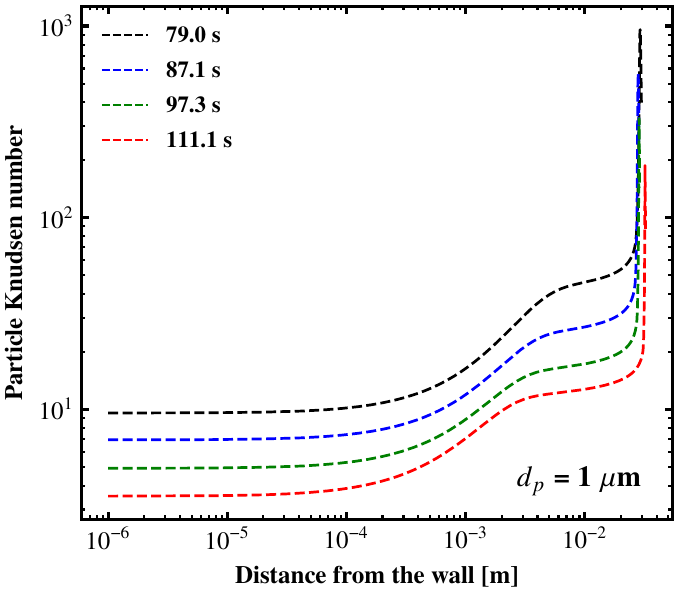}
    \caption{$d_p=1~\mu$m}
  \end{subfigure}
  \hfill
  \begin{subfigure}[t]{0.32\textwidth}
    \centering
    \includegraphics[width=\linewidth]{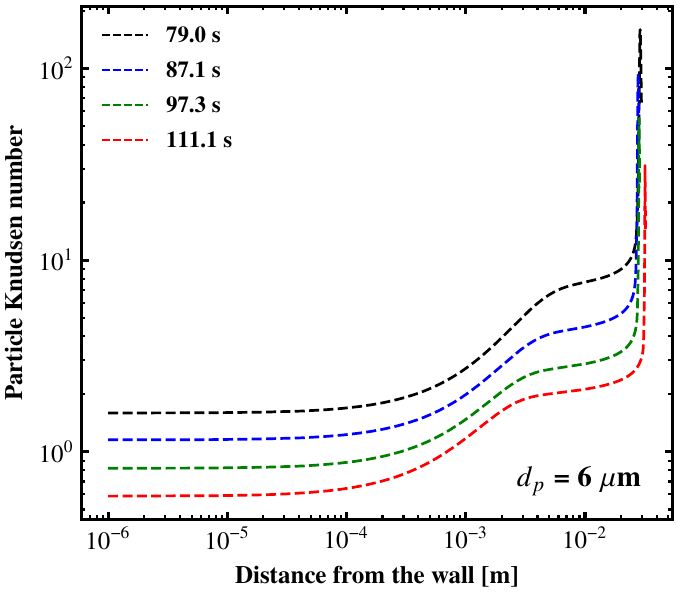}
    \caption{$d_p=6~\mu$m}
  \end{subfigure}
  \hfill
  \begin{subfigure}[t]{0.32\textwidth}
    \centering
    \includegraphics[width=\linewidth]{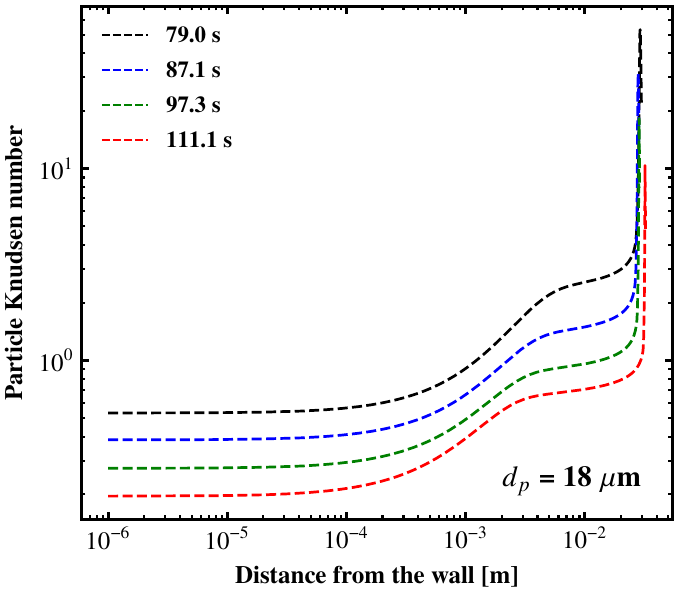}
    \caption{$d_p=18~\mu$m}
  \end{subfigure}
  \caption{Particle Knudsen number along the stagnation streamline at the four segment-end times for initial particle diameters of 1, 6, and 18~$\mu$m.}
  \label{fig:knp}
\end{figure}

The drag and heat transfer closures introduced above depend on the flow regime around each particle. Because the dust particles are only a few micrometers in diameter, this regime can be rarefied even where the flow around the vehicle is a continuum. As shown in Fig.~\ref{fig:knp}, the particle Knudsen number is therefore evaluated along the trajectory to identify the flow regime at the particle scale and to assess the applicability of these closures. It is defined as $Kn_p=\lambda/d_p$, with the local gas mean free path $\lambda$, and is related to the particle Mach and Reynolds numbers by $Kn_p=\sqrt{\gamma\pi/2}\,M_p/Re_p$, where $\gamma$ is the ratio of specific heats and $M_p$ and $Re_p$ are based on the slip velocity $|\boldsymbol{u}^g-\boldsymbol{v}_p|$. The particle conditions are evaluated with ORACLE along the stagnation streamline for initial diameters of 1, 6, and 18~$\mu$m, with $\lambda$ obtained from the hard-sphere relation $\lambda=(\mu^g/p^g)\sqrt{\pi R^g T_{\mathrm{tr}}^g/2}$, where $R^g$ is the specific gas constant of the mixture. Along the present trajectory, $\lambda$ in the shock layer ranges from approximately 3.5~$\mu$m near the wall to 110~$\mu$m behind the shock. Relative to the 0.6~m nose radius, the Knudsen number of the gas flow is of order $10^{-5}$ to $10^{-4}$, and the continuum formulation of Section~\ref{subsec:methods_noneq} applies. For particle diameters of 1--18~$\mu$m, however, $Kn_p$ ranges from approximately 0.2 to 100 as shown in Fig.~\ref{fig:knp}, placing the particles in the transitional and free-molecular regimes. In the shock layer, $M_p$ ranges from 2 to 6 and increases toward the wall, whereas $Re_p$ remains below approximately 40. These ranges include the particle conditions computed by Palmer and Sahai for the Schiaparelli shock layer, with $Re_p$ of 0.4--6, $M_p$ of 2--4, and $Kn_p$ of 0.8--10~\cite{palmer2023effect}.

The Henderson correlation covers these conditions, spanning the continuum, slip, transitional, and free-molecular regimes for Mach numbers up to 6~\cite{henderson1976drag}. For these conditions, Palmer and Sahai showed that a continuum drag correlation overestimates the particle deceleration, whereas the Henderson and Loth correlations predict similar surface recession~\cite{palmer2023effect}. The Fox correlation was proposed for Mach numbers up to 6 and Reynolds numbers up to $10^4$, and its agreement with experimental data was shown for $M_p/Re_p<0.5$~\cite{fox1978shock}. Most particles in the shock layer lie outside this range. The heat transfer closure, however, affects only the particle temperature and vaporization. The mechanical erosion in Eqs.~\eqref{eq:particle_penetration} and \eqref{eq:particle_erosion} depends on the impact velocity and particle diameter, and the collision heat flux in Eq.~\eqref{eq:particle_impact_energy} depends on the kinetic energy of the impacting particles. The impact velocity is determined by drag, which the Henderson correlation represents in the present regime. Particle vaporization is negligible in the present calculations, so the impact diameter remains equal to the initial diameter~\cite{palmer2020modeling,palmer2023effect,nam2026euler}. The predicted erosion and collision heating therefore depend on the drag closure rather than on the heat transfer closure.

Particles are injected at the inflow boundary using Monte Carlo sampling, with the injection rate determined by the particle mass loading. Momentum and heat exchange along the trajectories are accumulated within the gas phase cells using the particle-source-in-cell approach \cite{crowe1977particle}. The resulting source terms are transferred to HEGEL to account for two-way coupling.

Particle impacts contribute to surface heating and mechanical erosion. The kinetic energy lost by an impacting particle is assumed to be transferred to the wall as heat:
  \begin{equation}
    \Delta E_{p,k}
    =
    \frac{1}{2}m_{p,k}
    \left(
      |\boldsymbol{v}_{p,k}^{-}|^2
      -
      |\boldsymbol{v}_{p,k}^{+}|^2
    \right),
    \label{eq:particle_impact_energy}
  \end{equation}
where the superscripts $-$ and $+$ denote the states immediately before and after impact, respectively. The wall impact is treated as perfectly inelastic, so that $\boldsymbol{v}_{p,k}^{+}=0$ and the entire kinetic energy of the incident particle is converted into heat~\cite{nam2026euler}. The particle collision heat flux is obtained by summing the impact energy over all impacts, $q_{\mathrm{coll}}''=\sum_k \dot{N}_{A,k}\Delta E_{p,k}$, where $\dot{N}_{A,k}$ is the number of impacts per unit surface area and time for particle class $k$.

Mechanical erosion is evaluated using the crater-based correlation for Norcoat-Li\`ege used in
the Schiaparelli capsule~\cite{palmer2020modeling}. The penetration depth is expressed as
  \begin{equation}
    P_k
    =
    0.00028\,\rho_p^{0.62}d_{p,k}^{1.04867}
    |\boldsymbol{v}_{p,k}^{-}|^{0.667}.
    \label{eq:particle_penetration}
  \end{equation}
Assuming a hemispherical crater, the crater diameter is $D_{c,k}=2P_k$. The mechanical recession rate is then evaluated as
  \begin{equation}
    \dot{s}_{\mathrm{erosion}}
    \approx
    \sum_k
    \dot{N}_{A,k}D_{c,k}^{2}P_k\cos\theta,
    \label{eq:particle_erosion}
  \end{equation}
where $\dot{N}_{A,k}$ is the impact rate defined above. Following the adopted erosion model, an effective impact angle of $\theta=45^\circ$ is used
 \cite{palmer2020modeling,nam2026euler}.
The particle collision heat flux is passed to HEGEL, and the mechanical recession rate is passed to PATO (in-depth), allowing particle-induced heating and erosion to be included in the coupled material response.
A particle-flux-based trajectory control volume (TCV) method \cite{hinkle2024efficient} is adopted to smooth
the particle-induced surface quantities. 

\subsection{Material Response}\label{subsec:methods_material}

The in-depth material response is computed using the Porous material Analysis Toolbox based on OpenFOAM (PATO) \cite{meurisse2018multidimensional}. The material is treated as a porous medium consisting of a carbon-fiber phase, a phenolic matrix phase, and a gas phase. PATO solves the volume-averaged conservation equations for solid mass, gas mass, and energy, with pore-gas transport described by Darcy's law. The formulation accounts for heat conduction, solid decomposition, and energy transport by pyrolysis gases. Local thermal equilibrium is assumed between the solid and gas phases inside the material. The governing equations are described in detail by Meurisse et al.~\cite{meurisse2018multidimensional}. Within PATO, the thermodynamic properties of the pyrolysis gas are evaluated with the Mutation++ library~\cite{scoggins2020mutation}.

The material description follows the Theoretical Ablative Composite for Open Testing (TACOT) \cite{lachaud2014tacot}, a publicly available reference material representative of low-density carbon-phenolic ablators \cite{meurisse2018multidimensional}. Decomposition of the phenolic matrix is described by three Arrhenius reactions \cite{meurisse2018multidimensional}. For each reaction $j$, the advancement variable $\chi_j$ ($0 \leq \chi_j \leq 1$) is governed by
  \begin{equation}
  \frac{\partial \chi_j}{\partial t}
  =
  A_j T^{n_j}(1-\chi_j)^{m_j}
  \exp\left(-\frac{E_j}{R_u T}\right),
  \qquad j=1,2,3,
  \label{eq:material_pyrolysis}
  \end{equation}
where $A_j$ is the pre-exponential factor, $E_j$ is the activation energy, $R_u$ is the universal gas constant, and $n_j$ and $m_j$ are the temperature and reaction-order exponents, respectively. Table~\ref{tab:material_pyrolysis} summarizes the reaction parameters used in Eq.~\eqref{eq:material_pyrolysis}. The factor $F_j$ in Table~\ref{tab:material_pyrolysis} is the mass fraction of the matrix phase decomposed by reaction $j$, which weights the contribution of each reaction to the pyrolysis gas production rate~\cite{meurisse2018multidimensional}.

\begin{table}[htbp]
  \centering
  \caption{Arrhenius parameters for the TACOT matrix pyrolysis
  model \cite{meurisse2018multidimensional}.}
  \label{tab:material_pyrolysis}
  \begin{tabular}{cccccc}
  \toprule
  Reaction $j$ & $F_j$
  & $A_j$ [$\mathrm{s^{-1}}$]
  & $E_j$ [$\mathrm{J\,mol^{-1}}$]
  & $m_j$ & $n_j$ \\
  \midrule
  1 & 0.25 & $1.20\times10^{4}$ & $71130.89$ & 3 & 0 \\
  2 & 0.19 & $4.98\times10^{8}$ & $1.70\times10^{5}$ & 3 & 0 \\
  3 & 0.06 & $4.98\times10^{8}$ & $1.70\times10^{5}$ & 3 & 0 \\
  \bottomrule
  \end{tabular}
\end{table}

At the material surface, mass conservation accounts for the injection of char ablation products and pyrolysis gases with species diffusion in the gas phase. In elemental form, the surface mass balance is written as \cite{johnston2013influence}
\begin{equation}
  \dot{m}_c''(Z_{c,k}-Z_{w,k})
  +
  \dot{m}_g''(Z_{g,k}-Z_{w,k})
  -
  J_{k,n}
  =0,
  \label{eq:surface_mass_balance}
\end{equation}
where $\dot{m}_c''$ and $\dot{m}_g''$ are the char ablation and pyrolysis gas mass fluxes, respectively. The quantities $Z_{c,k}$, $Z_{g,k}$, and $Z_{w,k}$ denote the mass fractions of element $k$ in the char, pyrolysis gas, and wall gas. The quantity $J_{k,n}$ is the diffusive mass flux of element $k$ normal to the wall, directed into the gas. For the equilibrium model~\cite{meurisse2018multidimensional,le2025numerical}, the wall composition and char ablation rate are determined from the elemental balance and chemical equilibrium with solid carbon. For the finite-rate models, the corresponding species balances are closed using the surface reaction rates from Park-5 \cite{johnston2014study,le2025numerical} and the Zhluktov--Abe (ZA) model \cite{zhluktov1999viscous}.

The surface energy balance relates the applied heat flux to radiative cooling, mass-transfer enthalpy, and heat conduction into the material \cite{johnston2013influence, meurisse2018multidimensional}:
\begin{equation}
  \begin{split}
  q_{\mathrm{cond}}''
  ={}& q_{\mathrm{in}}''
  -\epsilon_w \sigma
  \left(T_w^4-T_{\infty}^4\right) \\
  &+\dot{m}_g''(h_g-h_w)
  +\dot{m}_c''(h_c-h_w),
  \end{split}
  \label{eq:surface_energy_balance}
\end{equation}
where $q_{\mathrm{in}}''$ is the heat flux supplied by HEGEL, including gas-side heat conduction and enthalpy transport by species diffusion. The conductive heat flux within the material, $q_{\mathrm{cond}}''$, is evaluated using Fourier's law. When particle impacts are considered, the particle collision heat flux is also included in $q_{\mathrm{in}}''$. The quantities $\epsilon_w$, $\sigma$, and  $T_{\infty}$ denote the surface emissivity, Stefan--Boltzmann constant, and freestream temperature, respectively. The specific enthalpies $h_g$,  $h_c$, and $h_w$ correspond to the pyrolysis gas, solid char,  and wall gas, respectively. The wall gas enthalpy is evaluated at the wall temperature and composition.

  \subsection{Coupling Strategy}
  \label{subsec:methods_coupling}

\begin{figure}[t]
      \centering
      \includegraphics[width=\textwidth]{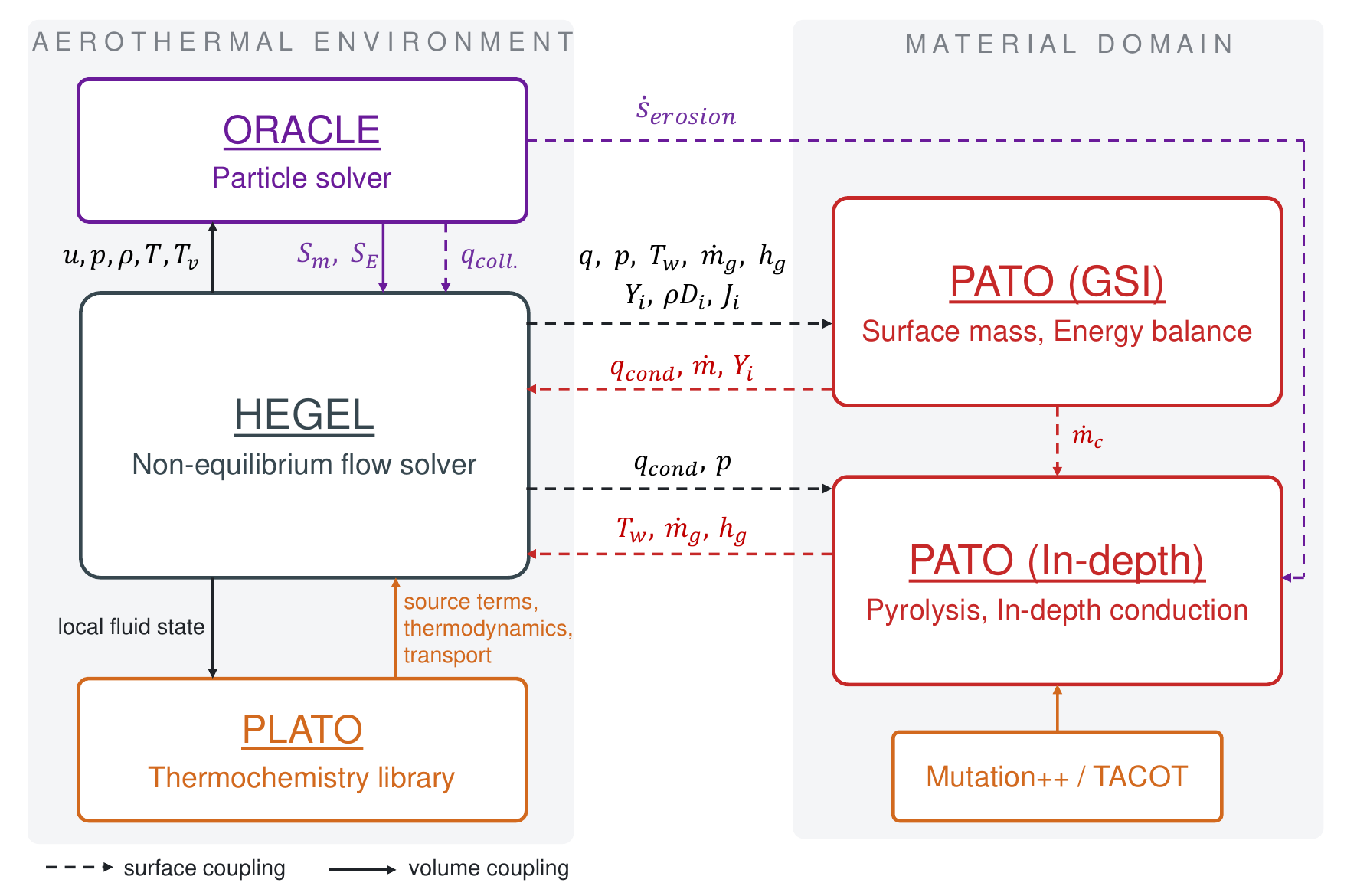}
      \caption{Schematic of the multiphysics coupling
      framework and exchanged quantities.}
      \label{fig:framework}
  \end{figure}

  HEGEL, ORACLE, and the two PATO participants exchange data through preCICE~\cite{bungartz2016precice}, as
  illustrated in Fig.~\ref{fig:framework}. The surface and in-depth material calculations are performed by separate PATO participants, denoted as PATO (GSI) and PATO (in-depth), respectively.  The coupled problem of the present study involves time scales that differ by several orders of magnitude. The gas flow and the particles cross the shock layer in less than a millisecond, whereas the in-depth material response evolves over the tens of seconds of the entry. The separate PATO participants allow each part of the problem to be advanced on its own time scale. PATO (in-depth) is coupled strongly with the flow, with implicit iterations over each physical time window $\Delta t_m$, so that the surface energy balance and the in-depth temperature remain consistent at the end of the window. Strong coupling is required at the gas-material interface because the surface mass and energy balances are stiff, with the blowing rate and heat fluxes varying strongly with wall temperature. Within each window, the surface balances in PATO (GSI) are solved without advancing physical time and are exchanged explicitly with HEGEL at every flow step. These $N$ flow-surface exchanges proceed in a pseudo-time with a step of $\Delta t_m/N$.

  PATO (in-depth) provides the wall temperature $T_w$, pyrolysis gas mass flux $\dot{m}_g''$, and gas enthalpy $h_g$. These quantities are transferred through HEGEL to PATO (GSI), together with the gas-side heat flux, pressure, and species transport quantities. PATO (GSI) evaluates the surface mass and energy balances to obtain the wall composition, char ablation mass flux, and conductive heat flux into the material. The wall composition and total blowing flux define the HEGEL wall boundary condition.
  The conductive heat flux $q_{\mathrm{cond}}''$ and wall pressure are transferred through HEGEL to PATO (in-depth). The char ablation flux is transferred directly from PATO (GSI) to PATO (in-depth) for the recession calculation, when enabled.

  For particle-laden calculations, HEGEL supplies ORACLE with the flow variables and thermophysical properties. ORACLE returns volumetric momentum and energy source terms and the particle collision heat flux to HEGEL. The collision heat flux contributes to the surface energy balance, whereas the mechanical erosion rate is transferred directly to PATO (in-depth). These particle quantities approach a quasi-steady state within each trajectory segment, and the particle phase is therefore coupled loosely, with explicit updates, for simplicity and computational efficiency. The two-way coupling keeps the particle source terms in the flow and the collision heat flux at the surface consistent with the same particle solution, and it allows the framework to be applied at higher mass loadings~\cite{ching2022simulations}. For the present dust profile, the particle source terms change the stagnation point heat flux by less than 4\%. Particle tracking is performed at prescribed coupling intervals, with the resulting source terms, collision heat flux, and erosion rate retained between updates.

  Within each material coupling iteration, the wall temperature, pyrolysis gas mass flux, and gas enthalpy
  supplied by PATO (in-depth) remain fixed while HEGEL and PATO (GSI) perform $N$ flow-surface exchanges.
  The flowfield, wall composition, and char ablation flux are updated at each exchange. 

  The material coupling is treated implicitly over each physical time window $\Delta t_m$, as summarized in
  Algorithm~\ref{alg:coupled_framework}. The material state is stored at the beginning of the window.
  PATO (in-depth) first computes a provisional material response over the window using the current interface data. After the subsequent $N$ flow-surface exchanges, the coupling convergence criteria are evaluated. If these criteria are not satisfied, the material state is restored to its checkpoint at the beginning of the window, and the same physical interval is recomputed using the updated interface data.
  The flowfield continues to evolve across HEGEL--PATO (GSI) coupling iterations without being restored to its state at the beginning of the window. Convergence is assessed using the relative coupling residuals of wall temperature and material conductive
  heat flux, and a criterion on the relative change in gas-side wall heat flux.
  Once these criteria are satisfied, the material state is accepted, and the calculation proceeds to the next physical time window.

  When recession is included, the surface velocity is obtained as $\dot{s}=\dot{m}_c''/\rho_s+\dot{s}_{\mathrm{erosion}}$, where $\rho_s$ is the local solid density at the surface computed by PATO (in-depth), and the blowing mass flux remains $\dot{m}_w''=\dot{m}_c''+\dot{m}_g''$. The material mesh follows the receding surface through Laplacian mesh motion, which is applied at each coupling iteration from the state at the beginning of the window. Surface quantities are transferred between the flow and material surface meshes with the nearest-neighbor mapping of preCICE, and the flow mesh adopts the receded surface at the start of each trajectory segment. The recession within a segment is less than 1.3~mm, which is small compared with the 0.6~m nose radius, so the flow geometry is held fixed within each segment.

\begin{algorithm}[htbp]
  \caption{Coupling procedure for flow, material response,
  and particles}
  \label{alg:coupled_framework}
  \begin{algorithmic}[1]
  \Require Initial states; physical window $\Delta t_m$;
           flow-surface exchanges $N$;
           convergence criteria; maximum iterations $K_{\max}$

  \State Initialize the physical time at $t_0$
  \While{$t<t_{\mathrm{end}}$}
    \State Store the material state at the beginning of the window
    \For{$k=1,\ldots,K_{\max}$}
      \If{$k>1$}
        \State Restore the material state; retain the flowfield
      \EndIf
      \State Advance PATO (in-depth) over $[t,t+\Delta t_m]$ 
      \State Supply the updated wall state to HEGEL and PATO (GSI)
      \For{$j=1,\ldots,N$}
        \State Advance HEGEL and update the PATO (GSI) surface solution 
        \State Exchange the updated flow and wall data
      \EndFor
      \State Run ORACLE at the prescribed interval to update particle source terms, collision heating, and erosion (if particle coupling is enabled)
      \State Transfer updated interface loads to PATO (in-depth)
      \State Evaluate the coupling convergence criteria
      \If{all criteria are satisfied}
        \State \textbf{break}
      \EndIf
    \EndFor
    \State Save the final state and flag the window if unconverged
    \State $t\gets t+\Delta t_m$
  \EndWhile
  \end{algorithmic}
  \end{algorithm}

\subsection{Validation of the Flow-Material Coupling Framework}
\label{subsec:methods_validation}

The flow-particle coupling was verified by Nam et al.~\cite{nam2026euler}. The validation here therefore focuses on the flow-material coupling, for which the graphite arc-jet case of Zibitsker et al.~\cite{zibitsker2023finite} provides surface temperature, in-depth temperature, and ablation rate data. The present results are compared with the NASA Ames temperature measurements and with the numerical predictions of Zibitsker et al.~\cite{zibitsker2023finite}.

The experiment used a 4-inch-diameter iso-$q$ graphite specimen tested in the NASA Ames Interaction Heating Facility and an air-argon test gas. The test environment had a total enthalpy of 20.7~MJ/kg and stagnation point heat flux and pressure of 13.06~MW/m$^2$ and 127.3~kPa, respectively, as measured with a hemispherical calorimeter~\cite{zibitsker2023finite}. In-depth specimen temperatures were measured using five thermocouples positioned at different depths. The computational inflow is the freestream state extracted upstream of the shock, as detailed in Ref.~\cite{zibitsker2023finite}.

The gas phase is modeled with 12 species: five air species (N$_2$, O$_2$, NO, N, O), argon, and six ablation-product species (C, C$_2$, C$_3$, CO, CO$_2$, CN), in thermochemical nonequilibrium. The gas phase reactions follow those of Zibitsker et al.~\cite{zibitsker2023finite}. Transport properties use the Blottner--Eucken--Wilke formulation, and species diffusion follows a modified Fick's law with a constant Lewis number of 1.4. Gas-surface interactions are modeled using the air-carbon ablation (ACA) finite-rate model of Prata et al.~\cite{prata2022air}.

The specimen is ultra-fine Poco graphite with a density of 1660~kg/m$^3$ and an initial temperature of 293.15~K. Temperature-dependent thermal conductivity, specific heat, and emissivity follow Zibitsker et al.~\cite{zibitsker2023finite}. The material response is modeled by heat conduction in the graphite coupon alone, with adiabatic conditions on unexposed boundaries. The geometry is assumed to be fixed, given the small measured stagnation point recession of approximately 0.1~mm~\cite{zibitsker2023finite}.
The heat flux supplied to the material is ramped up linearly over the first 0.35~s to represent specimen insertion. Coupled heating continues to 5.45~s, followed by a standalone PATO calculation with a linear heating reduction to 5.80~s and radiative cooling to 20~s.

\begin{figure}[t]
  \centering
  \includegraphics[width=0.7\textwidth]{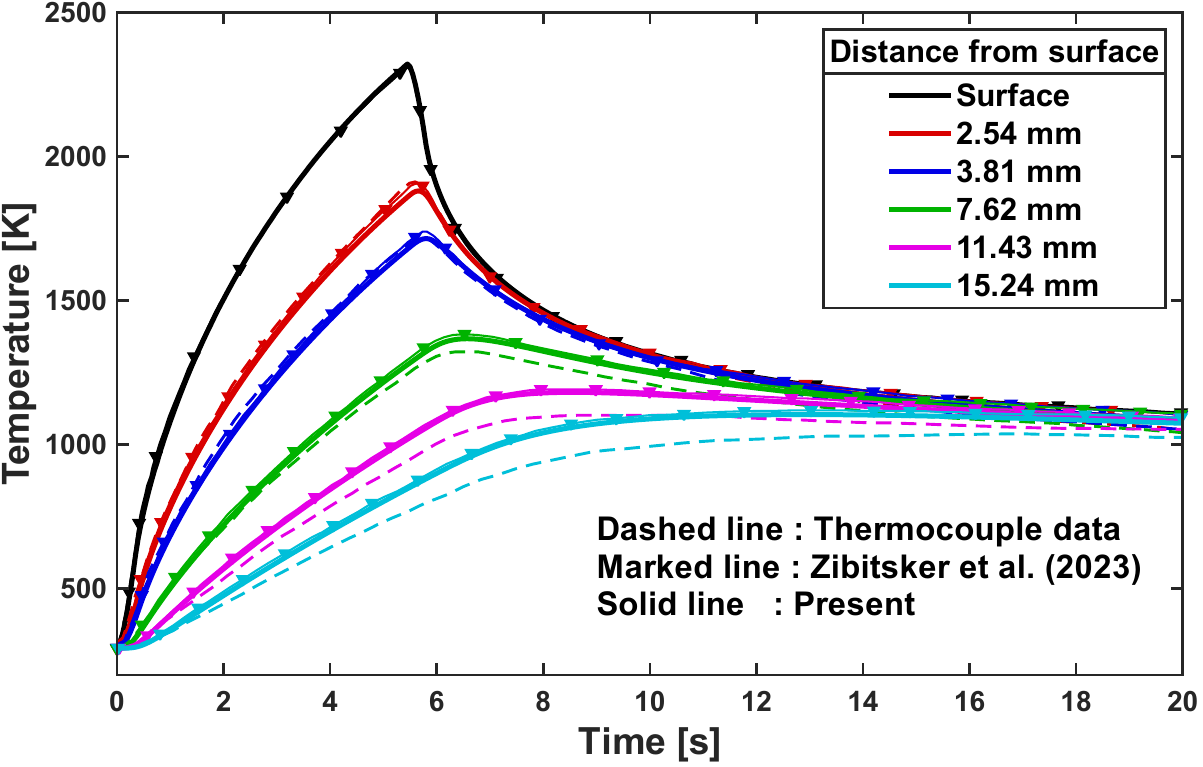}
  \caption{Comparison of in-depth temperature histories at the stagnation point for the iso-$q$ case.}
  \label{fig:isoq_temperature_history}
\end{figure}

Figure~\ref{fig:isoq_temperature_history} compares the surface temperature and the centerline temperatures at depths of 2.54, 3.81, 7.62, 11.43, and 15.24~mm. The present calculation captures the temperature rise during heating and the subsequent cooling of the specimen. 
The surface temperature reaches approximately 2300~K near the end of heating, while the deeper locations respond more slowly and reach their peaks later. The present results show good agreement with the numerical results of Zibitsker et al.~\cite{zibitsker2023finite}. The predicted temperatures agree well with measurements at depths of 2.54 and 3.81~mm but overestimate temperatures at deeper depths. Zibitsker et al.~\cite{zibitsker2023finite} also report this depth-dependent discrepancy and examine its sensitivity to material properties and transport modeling.

\begin{figure}[t]
  \centering
  \includegraphics[width=0.49\textwidth]{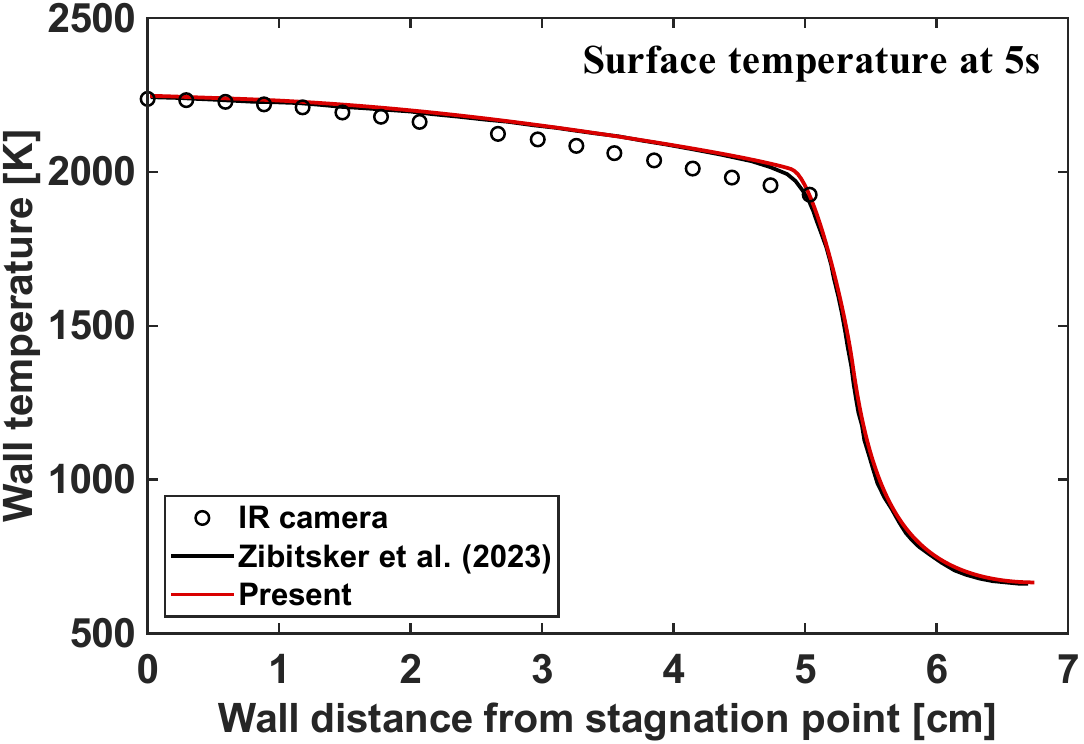}
  \includegraphics[width=0.49\textwidth]{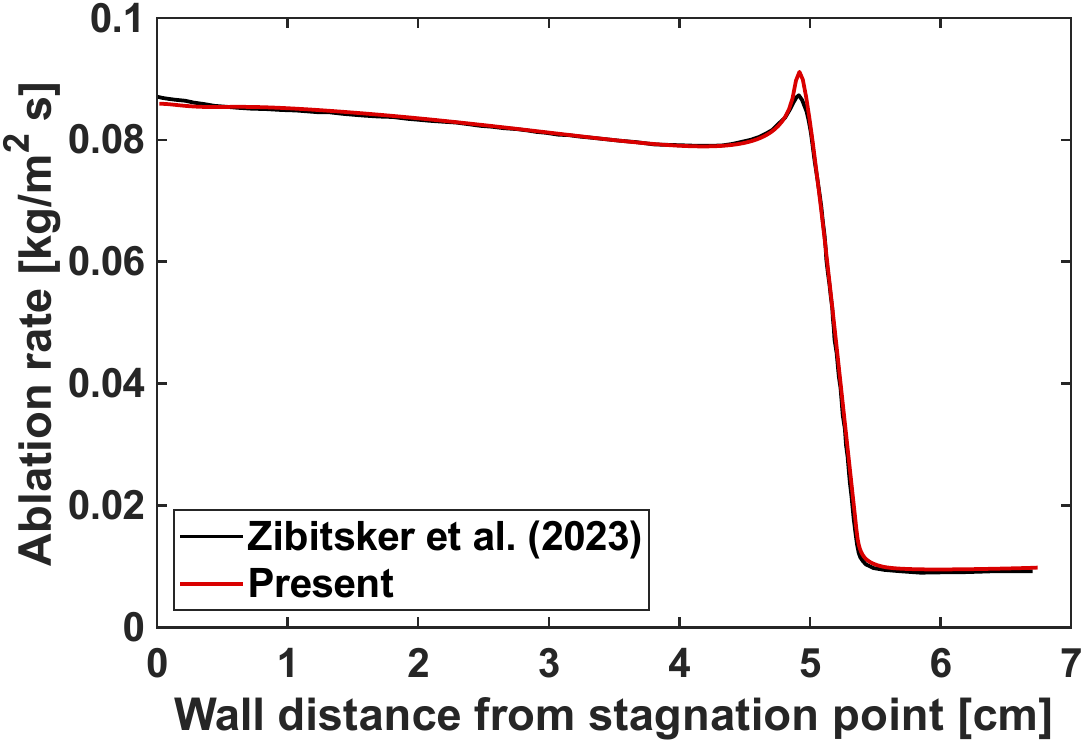}
  \caption{Surface distributions at $t=5$~s: temperature (left) and
  ablation mass flux (right) compared with the results of 
  Zibitsker et al.~\cite{zibitsker2023finite}.}
  \label{fig:isoq_surface_comparison}
\end{figure}

Figure~\ref{fig:isoq_surface_comparison} shows the surface temperature and ablation mass flux along the specimen surface at $t=5$~s.  The predicted surface temperature decreases from approximately 2240~K at the stagnation point to approximately 2000~K near the shoulder at 5~cm and then drops below 700~K along the side of the specimen. The prediction agrees with the IR camera measurements at the stagnation point and overestimates them by up to approximately 60~K between 3 and 5~cm. The ablation mass flux is approximately 0.086~kg/(m$^2$~s) at the stagnation point and decreases slightly toward the shoulder, where a local peak of approximately 0.09~kg/(m$^2$~s) appears. Beyond the shoulder, the flux decreases to approximately 0.01~kg/(m$^2$~s). Both distributions agree closely with the numerical predictions of Zibitsker et al.~\cite{zibitsker2023finite}, with the largest difference in the magnitude of the ablation peak at the shoulder. These comparisons show that the coupled flow-material framework reproduces the reference numerical solution and the measured surface temperature, which supports its use for the Mars entry calculations in Section~\ref{sec:results}.

\FloatBarrier
\section{Results and Discussion}\label{sec:results}

\subsection{Computational Conditions}
  \label{subsec:computational_conditions}

The Schiaparelli entry capsule entering the Martian atmosphere at hypersonic speeds is considered to investigate the combined effects of surface chemistry, geometric recession, and dust-particle impacts on the heat shield response. The simulations focus on the forebody, which has a sphere-cone geometry with a diameter of 2.4~m, a nose radius of 0.6~m, and a cone half-angle of $70^\circ$~\cite{palmer2020modeling,nam2026euler}. The computational domains comprise the gas surrounding the forebody and a 12.8-mm-thick heat shield~\cite{palmer2020modeling}, modeled using TACOT with a carbon-fiber phase and a phenolic resin matrix. The material properties of Norcoat-Li\`{e}ge required for the in-depth response and gas-surface interaction are not publicly available. The heat shield is therefore modeled with TACOT, and the Norcoat-Li\`{e}ge crater correlation, which was derived from arc-jet erosion tests for the Schiaparelli heat shield material~\cite{palmer2020modeling}, is applied to the receding TACOT surface.

The freestream conditions are taken from the Schiaparelli trajectory considered by Palmer et al.~\cite{palmer2020modeling}. Table~\ref{tab:trajectory} lists the selected trajectory points between 50 and 30~km, corresponding to flight times of approximately 71.7--111.1~s. This interval covers the peak-heating and peak particle erosion regions identified in the reference study, occurring near 45 and 35~km, respectively \cite{palmer2020modeling}. It is therefore selected to examine the combined effects of aerodynamic heating, thermochemical ablation, and particle-induced erosion. 

Four consecutive segments are computed: 50--45, 45--40, 40--35, and 35--30~km. The freestream conditions are held constant at the starting-point values within each segment. This piecewise-constant treatment approximates the continuous variation of the freestream conditions with a sequence of steady states and follows the approach used in previous trajectory analyses of particle-induced recession~\cite{nam2026euler,palmer2020modeling}. At each segment transition, the freestream conditions are updated, while the material temperature, decomposition state, and accumulated recession are inherited from the preceding calculation. This procedure preserves the material heating history across the trajectory segments. Results are reported by flight time, with segment-end comparisons at approximately 79.0, 87.1, 97.3, and 111.1~s.

\begin{table}[htbp]
  \centering
  \caption{Freestream conditions of the Schiaparelli entry trajectory.}
  \label{tab:trajectory}
  \small
\setlength{\tabcolsep}{4pt}
\begin{tabular}{cccccc}
    \toprule
    Altitude, & Flight time, & Density, & Velocity, & Temperature, & Mass loading, \\
    km & s & kg/m$^3$ & m/s & K & $\beta$ \\
    \midrule
    50.0 &  71.7 & $1.755\times10^{-4}$ & 5500.6 & 171.8 & $3.230\times10^{-5}$ \\
    45.0 &  79.0 & $2.944\times10^{-4}$ & 5185.0 & 175.0 & $4.140\times10^{-5}$ \\
    40.0 &  87.1 & $4.825\times10^{-4}$ & 4689.0 & 182.4 & $5.766\times10^{-5}$ \\
    35.0 &  97.3 & $7.717\times10^{-4}$ & 4016.9 & 186.3 & $6.283\times10^{-5}$ \\
    30.0 & 111.1 & $1.322\times10^{-3}$ & 2913.7 & 190.1 & $6.578\times10^{-5}$ \\
    \bottomrule
  \end{tabular}
\end{table}

The freestream gas is composed of 97\% CO$_2$ and 3\% N$_2$ by mass. The gas phase is modeled with 20 neutral species: N$_2$, O$_2$, N, NO, O, C, CN, CO, CO$_2$, C$_2$, C$_3$, CH, C$_2$H, CH$_2$, CH$_3$, CH$_4$, H, H$_2$, C$_2$H$_2$, and HCN. The char ablation products and pyrolysis gases injected from the TACOT surface are represented with the same species. This mixture and its gas phase reactions follow West and Johnston~\cite{west2025assessment}, with rate coefficients from Johnston~\cite{johnston2014study}. Transport properties are evaluated with the Chapman--Enskog method, using the collision integrals of Bellemans et al.~\cite{bellemans2019transport}, Laricchiuta et al.~\cite{laricchiuta2007classical}, and Bruno et al.~\cite{bruno2010transport}.

The flow domain is discretized with 200 cells normal to the wall and 300 cells along the wall, with a first-cell height of approximately 1~$\mu$m, and the grid is adapted to the bow shock~\cite{munafo2024hegel}. The heat shield is discretized with 50 cells through the thickness and 300 cells along the surface, with a first-cell size of 1~$\mu$m at the surface. The surface emissivity is interpolated between the TACOT virgin and char values of 0.8 and 0.9 according to the extent of pyrolysis. The back face and the shoulder boundary of the heat shield are adiabatic, and the material is initially at a uniform temperature of 300~K. The physical time window is $\Delta t_m=0.01$~s for the first segment and 0.003~s for the remaining segments, with $N=100$ flow-surface exchanges per window and $K_{\max}=200$. A window is accepted when the relative changes in wall temperature and conductive heat flux fall below $10^{-3}$ and $10^{-2}$, respectively, and the gas-side wall heat flux changes by less than $5\times10^{-3}$ over 20 flow steps. The particle solution is updated every 10 time windows.

The equilibrium~\cite{meurisse2018multidimensional,le2025numerical}, Park-5~\cite{johnston2014study,le2025numerical}, and Zhluktov--Abe~\cite{zhluktov1999viscous} surface chemistry models are each applied using the coupling procedure described in Section~\ref{subsec:methods_coupling}. The same prescribed trajectory and bulk material model are used for all three surface chemistry models. The flow solution accounts for the wall temperature, composition, and blowing supplied by the coupled surface and material calculations. The material response is advanced over physical time windows within each trajectory segment.

The Martian dust environment is based on the July 2007 global dust storm observed by the Mars Reconnaissance Orbiter, which represents a severe dust scenario for the Schiaparelli trajectory, following Palmer et al.~\cite{palmer2020modeling} and Nam et al.~\cite{nam2026euler}. The particle size distribution is represented by a modified gamma distribution, with particle radius ranging from 0.5 to 9~$\mu$m discretized into 15 size bins. The particle density is set to 2940~kg/m$^3$, and the mass loading, $\beta$, defined as the ratio of the particle mass per unit volume to the gas density, varies with altitude according to the prescribed dust profile~\cite{palmer2020modeling,nam2026euler}. Further details of the July 2007 dust environment modeling and verification of the HEGEL--ORACLE coupling framework are provided by Nam et al.~\cite{nam2026euler}.

Three configurations are considered for each surface chemistry model, resulting in nine combinations, as summarized in Table~\ref{tab:cases}. Case~1 uses a fixed surface geometry. Char ablation products and pyrolysis gases are injected into the flow, but char consumption is not converted into geometric recession. Case~2 includes surface recession due to char ablation, following the mesh motion procedure described in Section~\ref{subsec:methods_coupling}. Case~3 additionally includes particle effects, so that surface motion accounts for both char ablation and mechanical erosion. Particle-induced flow modification and collision heating are also retained in Case~3.

\begin{table}[htbp]
  \centering
  \caption{Definition of the three material configuration cases.}
  \label{tab:cases}
  \footnotesize
\setlength{\tabcolsep}{4pt}
\begin{tabular}{llccl}
    \toprule
    & Surface & Char & Particle & \\
    & motion  & recession & erosion & Description \\
    \midrule
    Case 1 & Fixed  & No  & No  & Surface chemistry without geometric recession \\
    Case 2 & Moving & Yes & No  & Char-driven recession \\
    Case 3 & Moving & Yes & Yes &  Char and particle-driven recession \\
    \bottomrule
  \end{tabular}
\end{table}

Cases~1 and 2 are compared to determine how char-induced surface recession affects the coupled response, while Cases~2 and 3 are compared to assess the additional contributions of particle heating and erosion. For each configuration, the three surface chemistry models are compared in terms of heat flux, wall temperature, mass injection, and cumulative recession.

  \subsection{Stagnation Point Analysis}\label{subsec:stag}

\begin{figure}[htbp]
  \centering
  \begin{subfigure}[t]{0.49\textwidth}
    \centering
    \includegraphics[width=\linewidth]{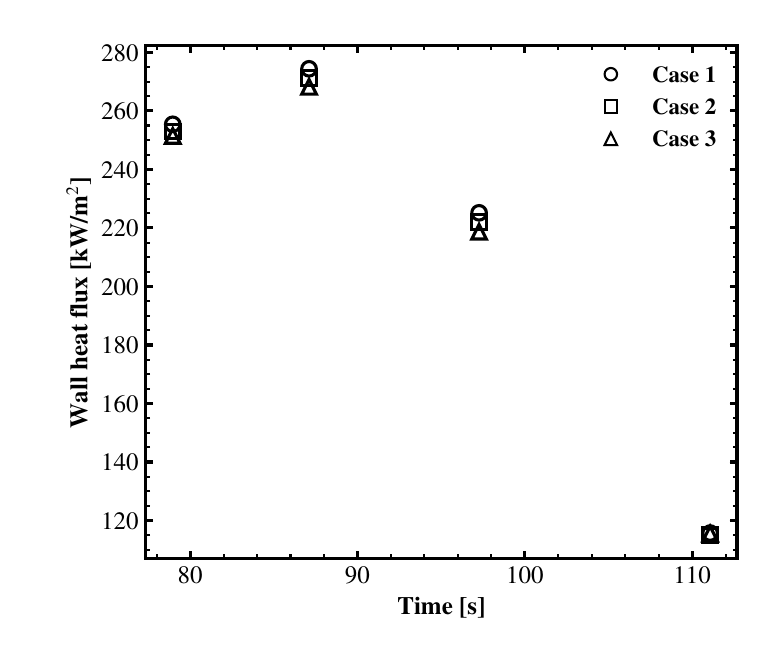}
    \caption{Wall heat flux}
    \label{fig:stag_eq_q}
  \end{subfigure}
  \hfill
  \begin{subfigure}[t]{0.49\textwidth}
    \centering
    \includegraphics[width=\linewidth]{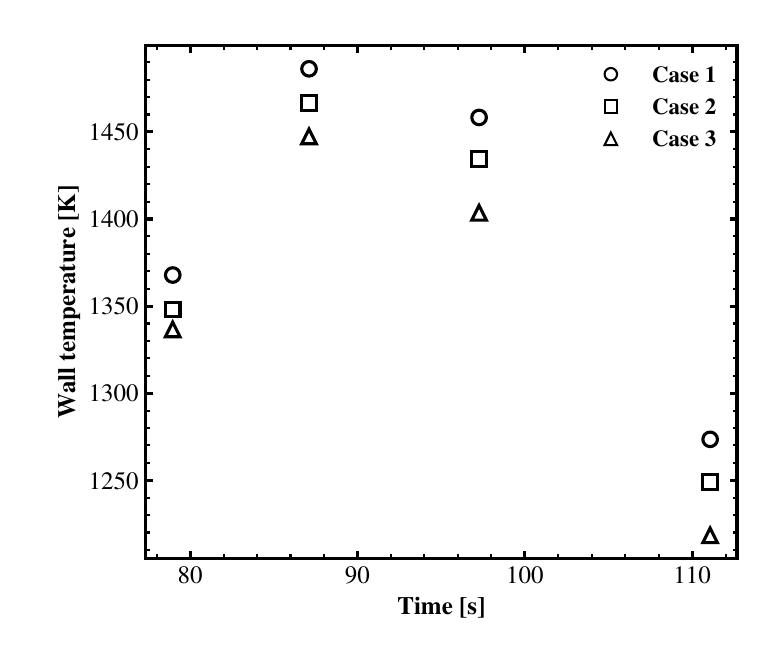}
    \caption{Wall temperature}
    \label{fig:stag_eq_T}
  \end{subfigure}
  \caption{Stagnation point values for the equilibrium  model, Cases~1--3.}
  \label{fig:stag_eq}
\end{figure}

The stagnation point response is compared at four segment-end times: 79.0, 87.1, 97.3, and 111.1~s. The results are shown as discrete points because each value is calculated at the end of a segment with prescribed, constant freestream conditions. These conditions are those at the start of each segment, so that the results at 111.1~s, for example, are obtained with the 35~km freestream conditions in Table~\ref{tab:trajectory}. Figure~\ref{fig:stag_eq} compares the equilibrium model predictions for the three configurations. Wall heat flux varies little among the cases, with differences below 1\% at 111.1~s, while wall pressure remains nearly unchanged. At 111.1~s, wall temperatures range from 1218 to 1273~K and decrease from Case~1 through Case~2 to Case~3. 

Cases~2 and 3 predict lower wall temperatures than Case~1, despite receiving similar gas-side heat fluxes. As the surface recedes, the heated char near the surface is removed, and material that was previously deeper and cooler forms the new surface. The temperature gradient near the surface therefore becomes steeper, and the reduced material thickness contributes to the same trend. By Fourier's law, this increases the heat conducted into the material, leaving less heat at the surface and resulting in a lower wall temperature. Case~3 includes the particle collision heating in the surface energy balance, which is about 1\% of the gas-side heat flux, yet it predicts the lowest wall temperature because particle erosion produces the largest recession.

\begin{figure}[htbp]
  \centering
  \begin{subfigure}[t]{0.49\textwidth}
    \centering
    \includegraphics[width=\linewidth]{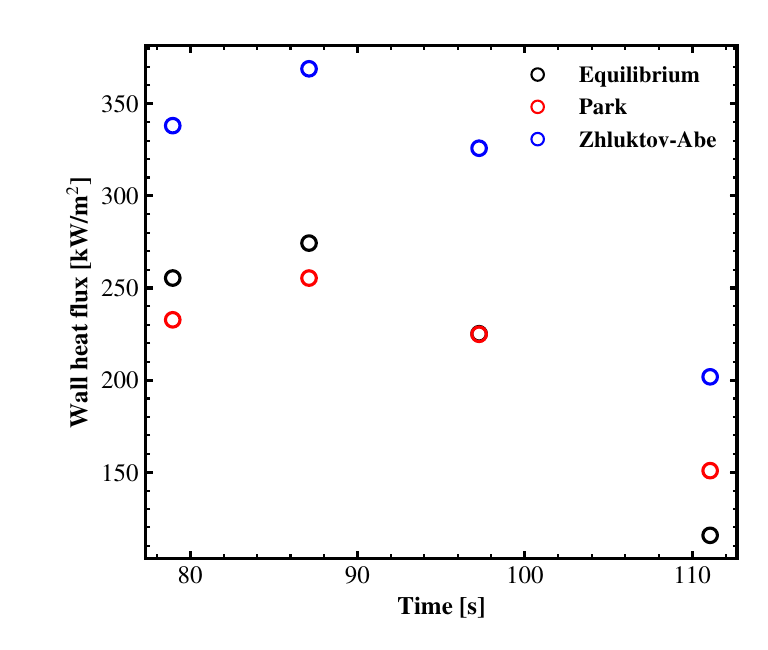}
    \caption{Wall heat flux}
    \label{fig:stag_mod_q}
  \end{subfigure}
  \hfill
  \begin{subfigure}[t]{0.49\textwidth}
    \centering
    \includegraphics[width=\linewidth]{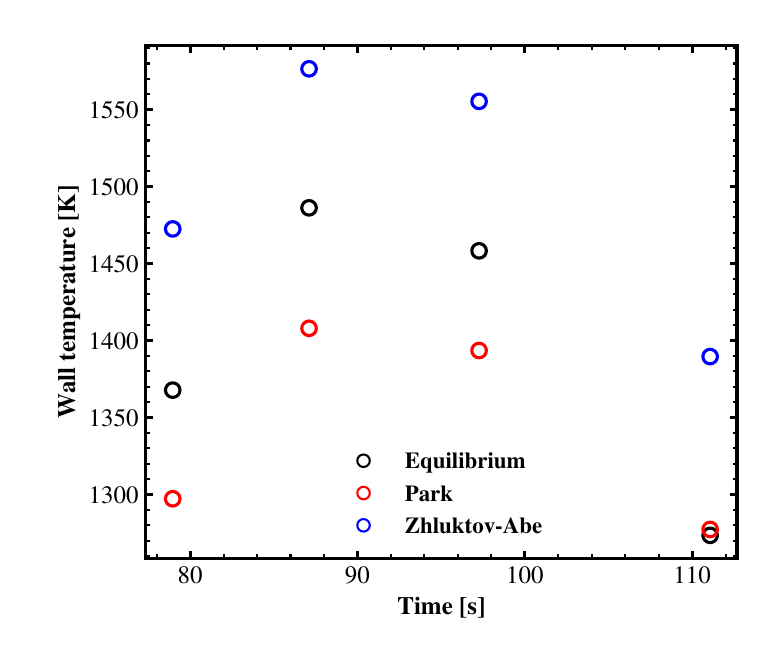}
    \caption{Wall temperature}
    \label{fig:stag_mod_T}
  \end{subfigure}
  \caption{Stagnation point values for the three surface chemistry models, Case~1.}
  \label{fig:stag_mod}
\end{figure}

Figure~\ref{fig:stag_mod} compares the surface chemistry models for Case~1. At each of the four times, the ZA model predicts the highest wall heat flux and temperature. The relative ordering of the Park-5 and equilibrium models changes over the trajectory: at 79.0~s, Park-5 predicts approximately 9\% lower heat flux, whereas at 111.1~s, it predicts 30\% higher heat flux. The wall temperature ordering also reverses, though the final temperatures are similar.

The change in heat flux ordering is accompanied by different trends in char injection. During the later part of the trajectory, the Park-5 char mass flux decreases by approximately 60\%, while the equilibrium model maintains a higher char flux. This greater mass injection in the equilibrium calculation contributes to its lower heat flux. Changes in wall composition and species diffusion enthalpy transport also affect the comparison. The ZA model predicts negligible char injection under the present conditions and thus produces the highest heat flux. Overall, differences in heat flux between surface chemistry models are substantially larger than differences among the three configurations for the equilibrium model.

\subsection{Surface Recession}

\begin{figure}[htbp]
  \centering
  \begin{subfigure}[t]{0.49\textwidth}
    \centering
    \includegraphics[width=\linewidth]{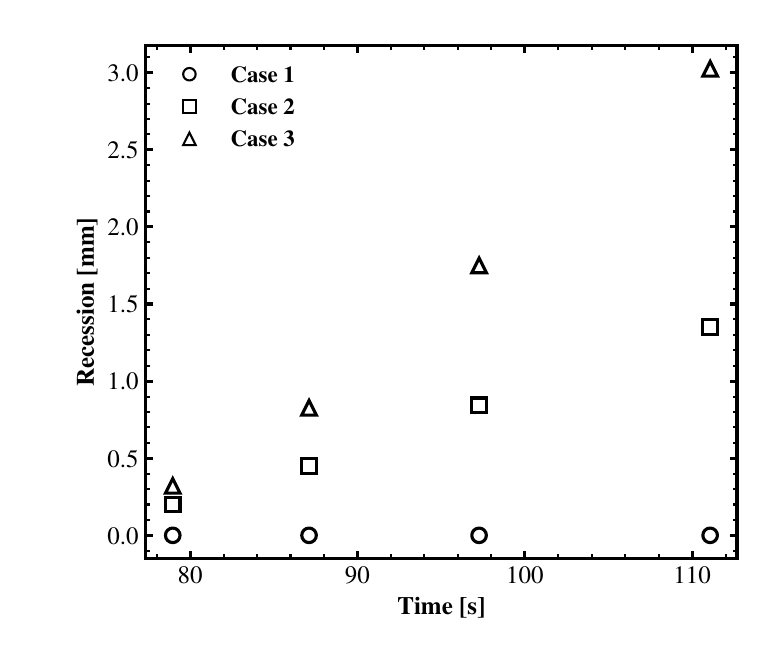}
    \caption{Equilibrium model, Cases~1--3}
    \label{fig:rec_eq}
  \end{subfigure}
  \hfill
  \begin{subfigure}[t]{0.49\textwidth}
    \centering
    \includegraphics[width=\linewidth]{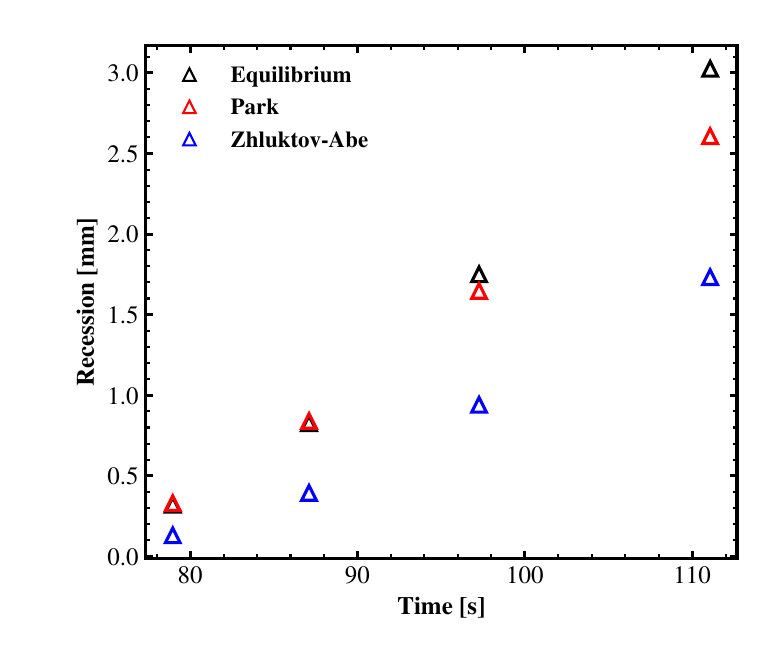}
    \caption{Three surface chemistry models, Case~3}
    \label{fig:rec_mod}
  \end{subfigure}
  \caption{Cumulative stagnation point recession.}
  \label{fig:rec}
\end{figure}

Figure~\ref{fig:rec} compares the cumulative stagnation point recession. At 111.1~s, including particle effects more than doubles the final recession for all three models. Case~2 reaches 1.35, 0.89, and approximately 0~mm for the equilibrium, Park-5, and ZA models, respectively, and Case~3 reaches 3.03, 2.61, and 1.73~mm. Case~1 retains its initial geometry, even though the flow calculation accounts for char injection. The Case~3 recession can be divided into char ablation and particle erosion contributions. The particle contributions are 1.731, 1.731, and 1.732~mm for the equilibrium, Park-5, and ZA models, respectively, showing a negligible effect of the surface chemistry model on the particle-induced surface recession. The corresponding char contributions are 1.294, 0.876, and approximately 0~mm.
The near-zero char contribution of the ZA model follows from its oxidation mechanism (Appendix~\ref{app:surface}). In Park-5, incident atomic oxygen removes carbon directly as CO (Table~\ref{tab:park5}). In ZA, carbon is removed mainly by atomic oxygen reacting with adsorbed oxygen, O(s), because the reactions of adsorbed oxygen with carbon alone have activation temperatures of 40{,}000~K and are slow at the present wall temperatures (Table~\ref{tab:za}). The ZA char flux therefore scales with the oxygen coverage $\Theta_{\mathrm{O}}$. Near the wall, the gas consists mostly of CO$_2$ and CO, and the exchange reaction CO$_2$ + (s) $\rightleftharpoons$ O(s) + CO, which accounts for more than 99\% of the O(s) formation and consumption, keeps $\Theta_{\mathrm{O}}$ at approximately $10^{-3}$. At the stagnation point at 87.1~s, the ZA char mass flux is $3.2\times10^{-6}$~kg/(m$^2$~s), approximately 1/1000 of the Park-5 oxidation rate evaluated at the same wall state.
 For all three models, mechanical erosion exceeds char-induced recession over the simulated time.

The nearly equal erosion contributions show that differences in surface chemistry have little effect on accumulated particle erosion in these calculations, despite the stagnation point blowing rate differing by a factor of 1.3 to 1.8 among the three models. By contrast, the char contribution changes substantially, driving the differences in total recession among the models.

\begin{figure}[htbp]
  \centering
  \begin{subfigure}[t]{0.49\textwidth}
    \centering
    \includegraphics[width=\linewidth]{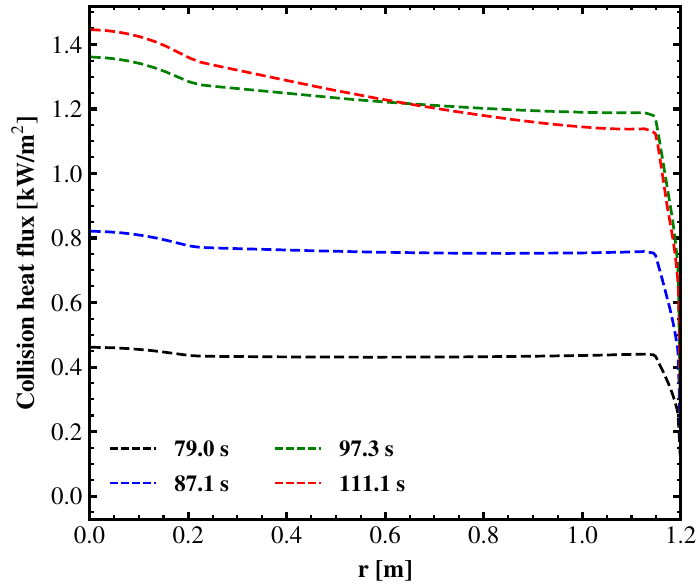}
    \caption{Collision heat flux}
  \end{subfigure}
  \hfill
  \begin{subfigure}[t]{0.49\textwidth}
    \centering
    \includegraphics[width=\linewidth]{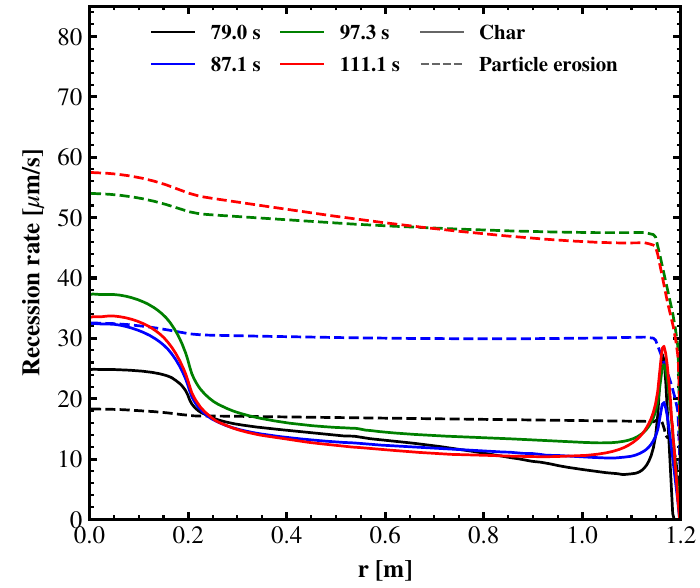}
    \caption{Recession rates}
  \end{subfigure}

  \vspace{1ex}
  \begin{subfigure}[t]{0.49\textwidth}
    \centering
    \includegraphics[width=\linewidth]{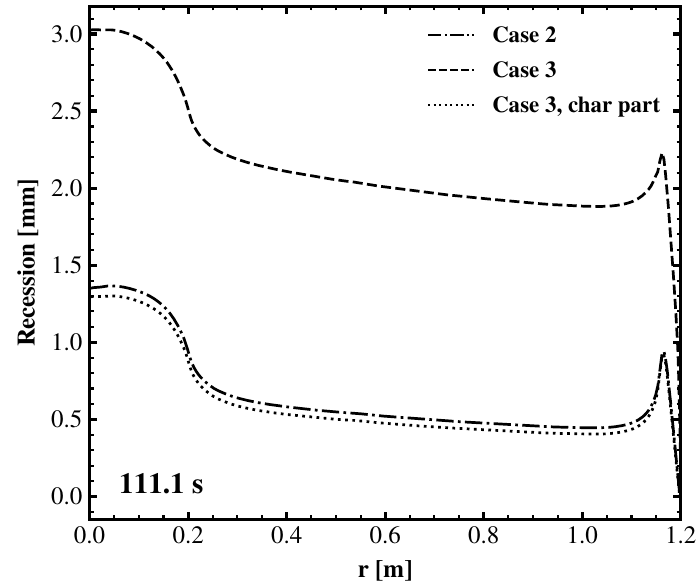}
    \caption{Recession at 111.1~s}
  \end{subfigure}
  \hfill
  \begin{subfigure}[t]{0.49\textwidth}
    \centering
    \includegraphics[width=\linewidth]{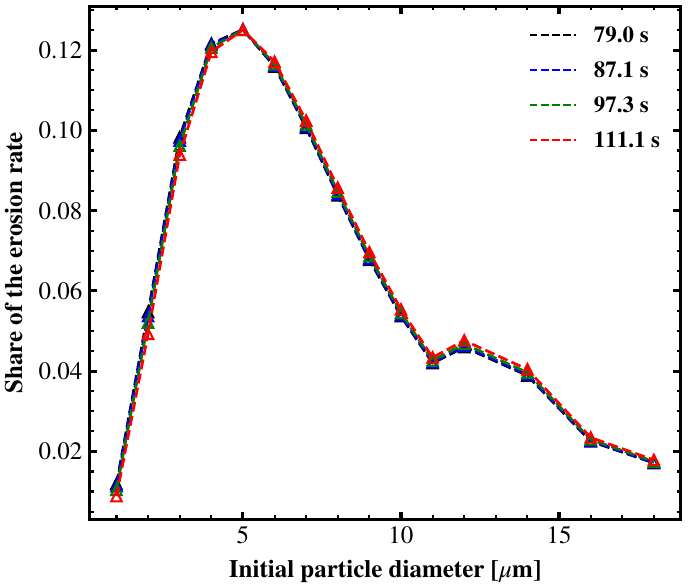}
    \caption{Erosion share by particle size}
  \end{subfigure}
  \caption{Particle-surface quantities for the equilibrium model, Case~3: (a) collision heat flux, (b) char recession and particle erosion rates, (c) recession profiles at 111.1~s for Cases~2 and 3, and (d) share of the stagnation point erosion rate by initial particle diameter.}
  \label{fig:particle_surface}
\end{figure}

Figure~\ref{fig:particle_surface} shows the distributions of the particle-surface quantities for the equilibrium model in Case~3. The collision heat flux at the stagnation point increases from approximately 0.46 to 1.45~kW/m$^2$ along the trajectory, which is about 1\% of the gas-side heat flux, and it is nearly uniform over the forebody (Fig.~\ref{fig:particle_surface}a). The erosion rate is also nearly uniform, whereas the char recession rate is concentrated near the nose (Fig.~\ref{fig:particle_surface}b). Erosion therefore exceeds char recession over the flank throughout the trajectory. At the stagnation point, it exceeds char recession from 97.3~s onward, reaching 57~$\mu$m/s at 111.1~s compared with 33~$\mu$m/s for char recession. In the recession profiles at 111.1~s, particle erosion adds a nearly uniform recession of approximately 1.5--1.7~mm over the forebody to the char-induced recession (Fig.~\ref{fig:particle_surface}c). Particles of 4--6~$\mu$m diameter contribute most to the erosion rate, and this size dependence remains nearly unchanged along the trajectory (Fig.~\ref{fig:particle_surface}d).

\begin{figure}[htbp]
  \centering
  \includegraphics[width=0.8\textwidth]%
    {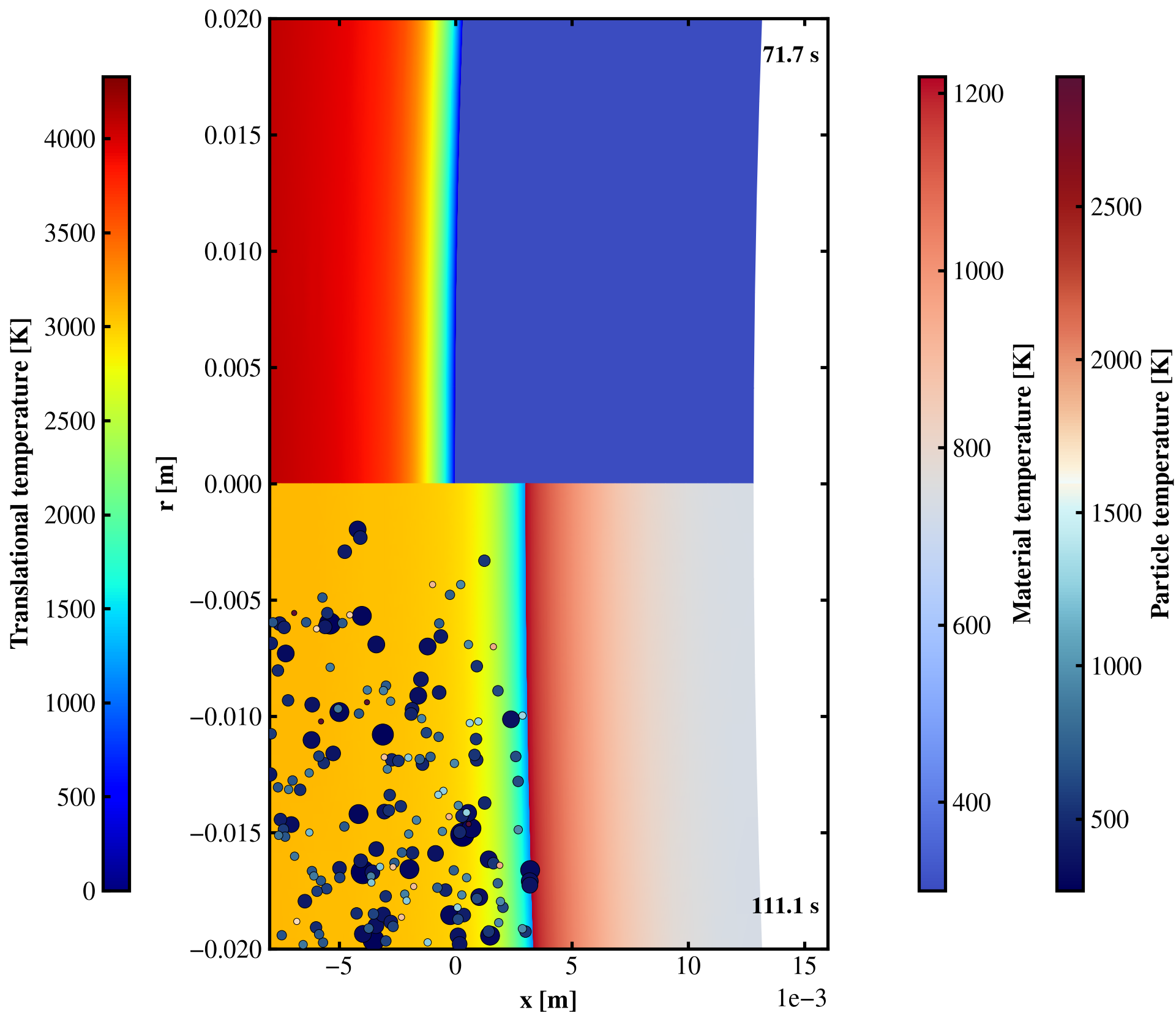}
  \caption{Nose region for the equilibrium model, Case 3, at 71.7~s (upper) and 111.1~s (lower), with the gas translational temperature, the material temperature, and the particle temperature.}
  \label{fig:nose_recession_case3}
\end{figure}

Figure~\ref{fig:nose_recession_case3} shows the nose geometry and temperature fields for the equilibrium model in Case~3 at 71.7 and 111.1~s. The shift between the initial and final surfaces represents approximately 3.03~mm of recession, which is approximately 24\% of the 12.8~mm thickness of the heat shield. During this interval, the material, initially at 300~K, develops a temperature gradient, reaching above 1200~K near the exposed surface and decreasing toward the interior. In the lower panel, the smaller particles are heated to approximately 2000--2500~K in the shock layer, whereas the largest particles remain below approximately 500~K before impact.

\subsection{Rarefaction Effects on Particle Erosion}\label{subsec:rarefaction}

Previous studies compared the particle-induced recession among drag correlations that include rarefaction and compressibility effects and found similar results for the Henderson, Loth, and Singh correlations~\cite{palmer2023effect,ching2022simulations,kroells2025investigation}. Palmer and Sahai~\cite{palmer2023effect} also showed that neglecting rarefaction and compressibility substantially overestimates the drag coefficient and the deceleration of dust particles in the Schiaparelli shock layer. To the authors' knowledge, the resulting change in the erosion has not been reported. Here, the particles are re-tracked on the equilibrium model, Case~3 flowfields at 87.1 and 111.1~s with the Clift--Gauvin correlation~\cite{clift1971motion}, which represents continuum, incompressible drag and therefore neglects both rarefaction and compressibility. All other models and the particle seeding are unchanged. Because the particle source terms have a small effect on the flow, the re-tracking is performed on fixed flowfields.

\begin{figure}[htbp]
  \centering
  \begin{subfigure}[t]{0.49\textwidth}
    \centering
    \includegraphics[width=\linewidth]{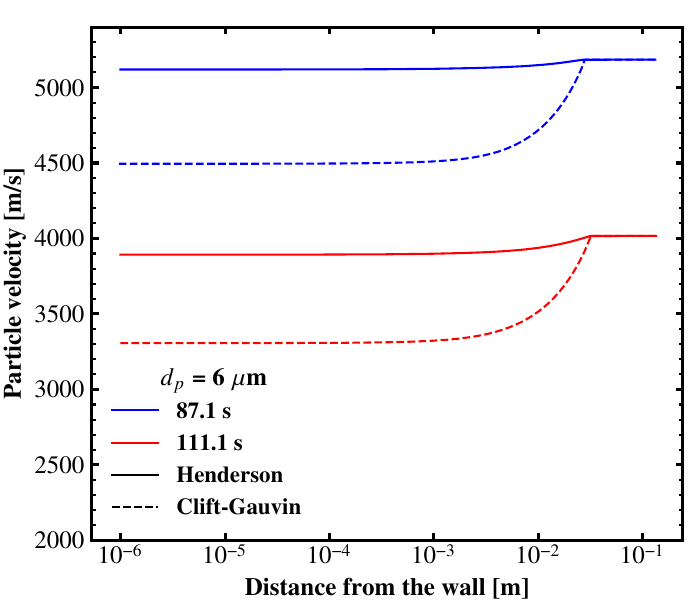}
    \caption{Particle velocity, $d_p=6~\mu$m}
  \end{subfigure}
  \hfill
  \begin{subfigure}[t]{0.49\textwidth}
    \centering
    \includegraphics[width=\linewidth]{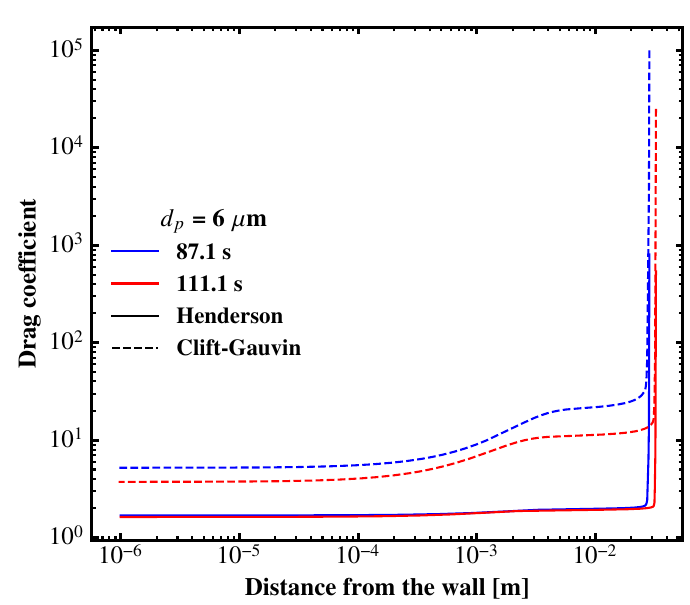}
    \caption{Drag coefficient, $d_p=6~\mu$m}
  \end{subfigure}

  \vspace{1ex}
  \begin{subfigure}[t]{0.49\textwidth}
    \centering
    \includegraphics[width=\linewidth]{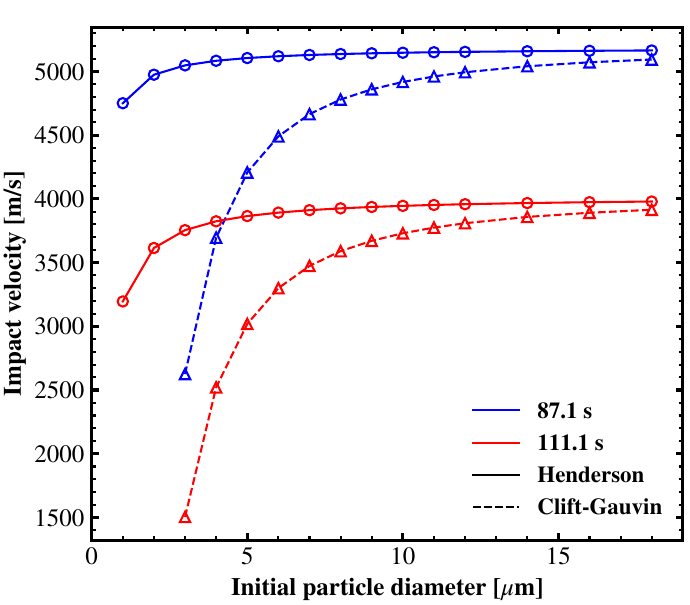}
    \caption{Impact velocity}
  \end{subfigure}
  \hfill
  \begin{subfigure}[t]{0.49\textwidth}
    \centering
    \includegraphics[width=\linewidth]{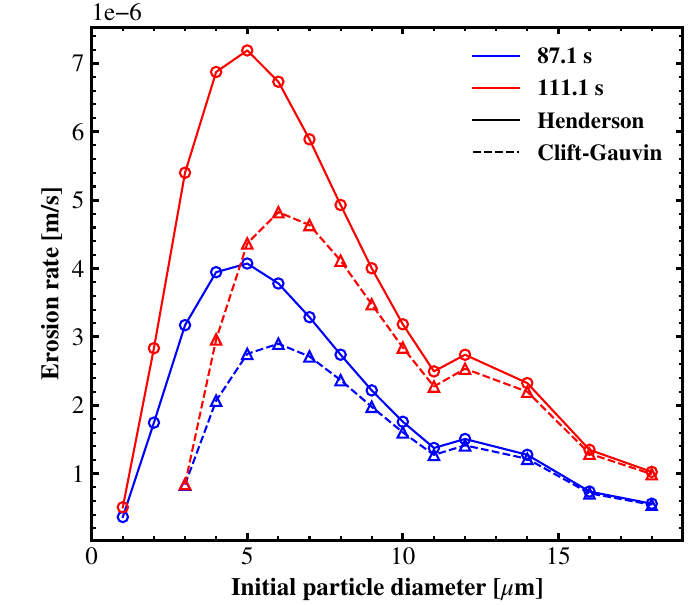}
    \caption{Stagnation point erosion rate}
  \end{subfigure}
  \caption{Comparison of the Henderson and Clift--Gauvin drag correlations for the equilibrium model, Case~3, at 87.1 and 111.1~s: (a) particle velocity and (b) drag coefficient along the stagnation streamline for $d_p=6~\mu$m, and (c) impact velocity and (d) stagnation point erosion rate by initial particle diameter.}
  \label{fig:drag_cmp}
\end{figure}

Figure~\ref{fig:drag_cmp} compares the results with those of the Henderson correlation. For 6~$\mu$m particles at particle Knudsen numbers of approximately 1--2, the continuum correlation predicts drag coefficients 2 to 5 times higher than the Henderson correlation (Fig.~\ref{fig:drag_cmp}b), and the particles decelerate more strongly in the shock layer (Fig.~\ref{fig:drag_cmp}a). The impact velocity decreases most for small particles. At 111.1~s, 3~$\mu$m particles impact at approximately 1500~m/s instead of 3760~m/s, and 1 and 2~$\mu$m particles do not reach the stagnation point (Fig.~\ref{fig:drag_cmp}c). Particles larger than about 10~$\mu$m are less affected, with impact velocities within 6\% of the Henderson values. Because particles of 3--7~$\mu$m dominate the erosion rate, the stagnation point erosion rate decreases by 31\% at 87.1~s and 35\% at 111.1~s (Fig.~\ref{fig:drag_cmp}d), which corresponds to 0.08 and 0.28~mm less recession over the respective segments. Neglecting rarefaction and compressibility at the particle scale therefore underestimates the particle-induced recession by approximately one third, which supports the use of a drag closure that covers the transitional and free-molecular regimes.

\subsection{Evolution of the Surface Distributions}

\begin{figure}[htbp]
  \centering
  \begin{subfigure}[t]{0.49\textwidth}
    \centering
    \includegraphics[width=\linewidth]{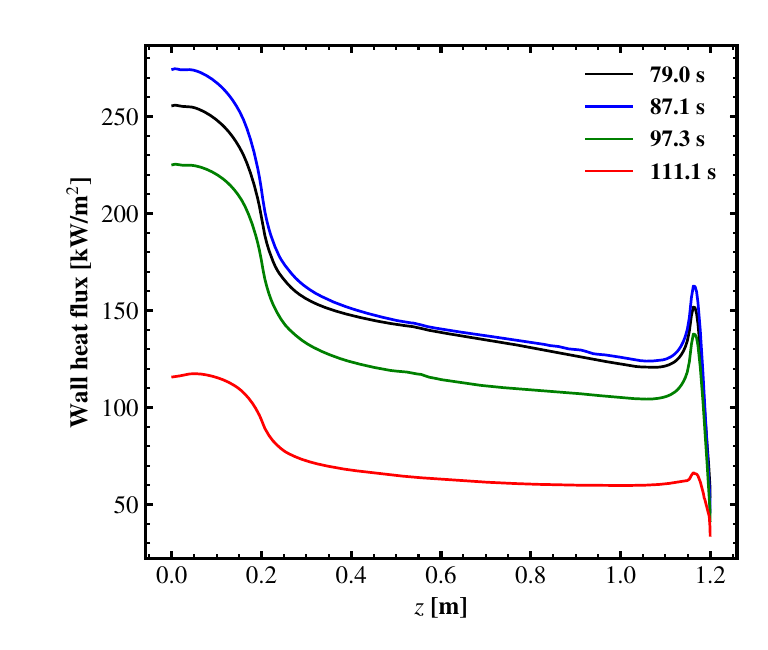}
    \caption{Wall heat flux}
    \label{fig:surfT_c1_qw}
  \end{subfigure}
  \hfill
  \begin{subfigure}[t]{0.49\textwidth}
    \centering
    \includegraphics[width=\linewidth]{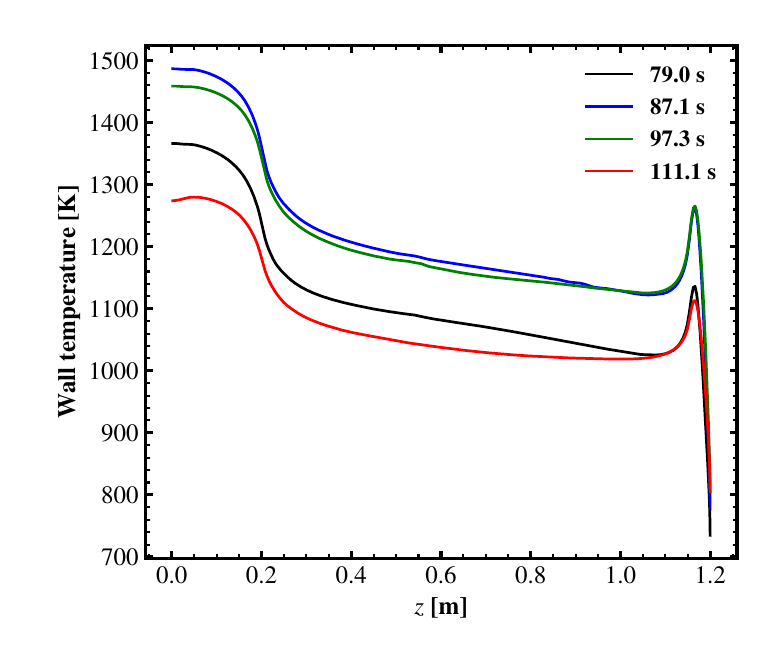}
    \caption{Wall temperature}
    \label{fig:surfT_c1_Tw}
  \end{subfigure}

  \vspace{1ex}
  \begin{subfigure}[t]{0.49\textwidth}
    \centering
    \includegraphics[width=\linewidth]{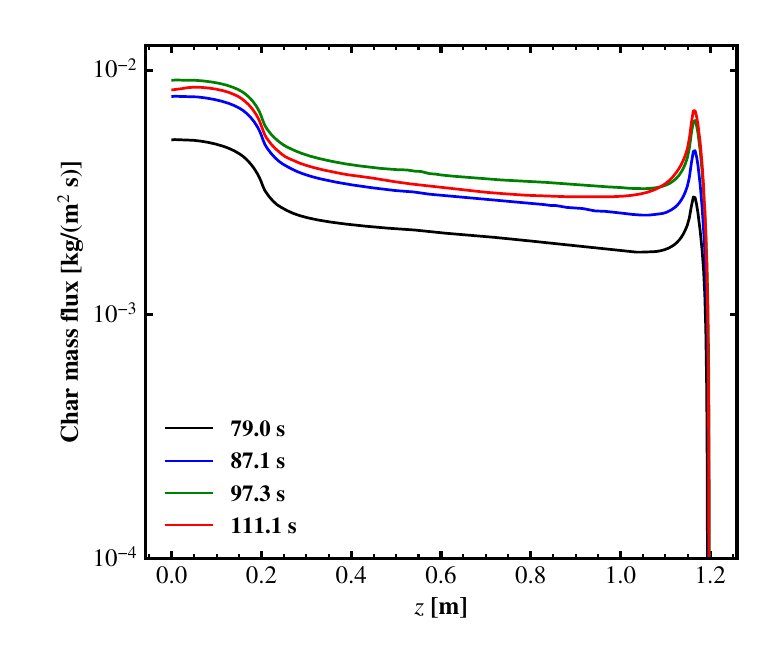}
    \caption{Char mass flux}
    \label{fig:surfT_c1_mdotC}
  \end{subfigure}
  \hfill
  \begin{subfigure}[t]{0.49\textwidth}
    \centering
    \includegraphics[width=\linewidth]{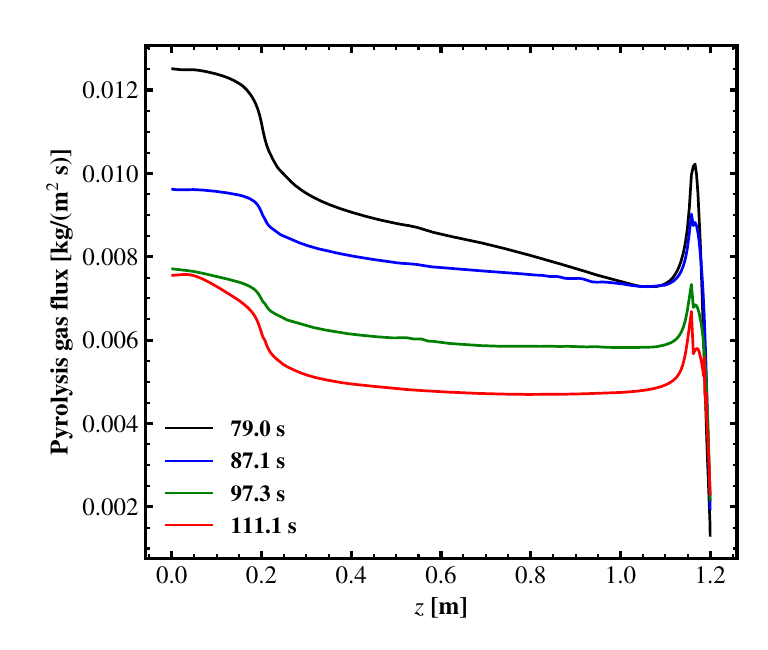}
    \caption{Pyrolysis gas flux}
    \label{fig:surfT_c1_mdotG}
  \end{subfigure}
  \caption{Wall distributions at the four segment-end times for the equilibrium  model, Case~1.}
  \label{fig:surfT_eq_case1}
\end{figure}

Figure~\ref{fig:surfT_eq_case1} shows the surface distributions for Case~1 at the four segment-end times. The profiles retain a similar overall shape, with high values near the nose and a local increase near the shoulder, where the boundary layer thins as the flow turns around the shoulder. Among the sampled times, the nose heat flux and wall temperature are highest at 87.1~s and decrease thereafter.

The pyrolysis gas flux follows a different trend, decreasing over the forebody across the four segment-end times. Pyrolysis gas is generated at a front that advances into the material, leaving char behind~\cite{ewing2013numerical}. Its evolution is therefore governed by the material’s internal state rather than by surface heating alone. As decomposition progresses, less resin remains in the heated layers, and the front moves deeper into the material. Heat must then travel farther to reach undecomposed resin. These effects reduce the resin decomposition rate, causing the gas flux to decrease even as surface heating passes through its maximum.

Unlike the heat flux and the pyrolysis gas flux, the char mass flux continues to increase after the heating peak (Fig.~\ref{fig:surfT_eq_case1}c). At the stagnation point, it increases from approximately $5.2\times10^{-3}$~kg/(m$^2$~s) at 79.0~s to $9.1\times10^{-3}$~kg/(m$^2$~s) at 97.3~s and decreases slightly to $8.3\times10^{-3}$~kg/(m$^2$~s) at 111.1~s. The char flux follows the supply of oxidizing species to the surface, which increases with the freestream density as the vehicle descends, whereas the convective heating decreases with the freestream velocity after 87.1~s. Because the char injection increases while the pyrolysis gas flux decreases, the total blowing flux at the stagnation point decreases by only about 10\% over the trajectory, from 0.0177 to 0.0159~kg/(m$^2$~s). For the Park-5 model, the char flux instead reaches its maximum at 87.1~s and decreases by approximately 60\% by 111.1~s, following the decrease in wall temperature.

\subsection{Surface Distributions at 111.1~s}

\begin{figure}[htbp]
  \centering
  \begin{subfigure}[t]{0.49\textwidth}
    \centering
    \includegraphics[width=\linewidth]{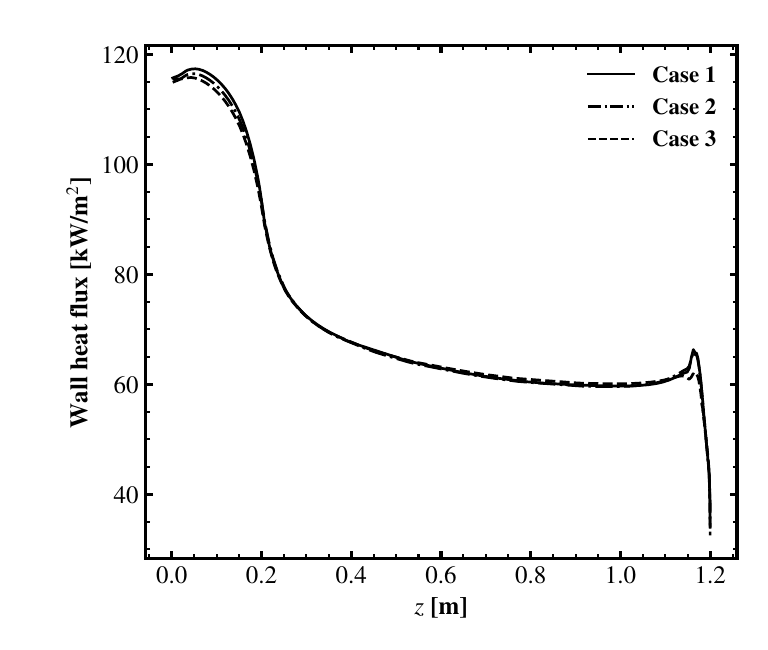}
    \caption{Wall heat flux}
  \end{subfigure}
  \hfill
  \begin{subfigure}[t]{0.49\textwidth}
    \centering
    \includegraphics[width=\linewidth]{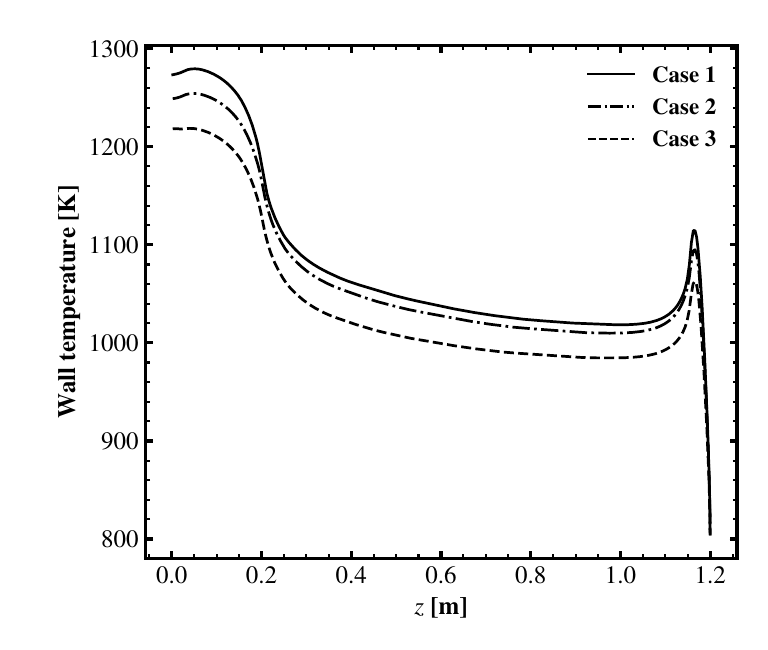}
    \caption{Wall temperature}
  \end{subfigure}

  \vspace{1ex}
  \begin{subfigure}[t]{0.49\textwidth}
    \centering
    \includegraphics[width=\linewidth]{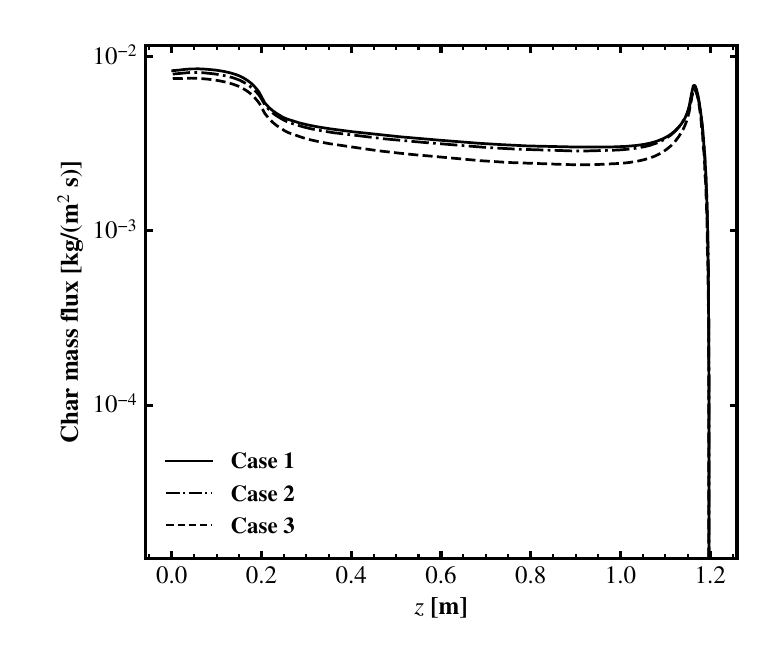}
    \caption{Char mass flux}
  \end{subfigure}
  \hfill
  \begin{subfigure}[t]{0.49\textwidth}
    \centering
    \includegraphics[width=\linewidth]{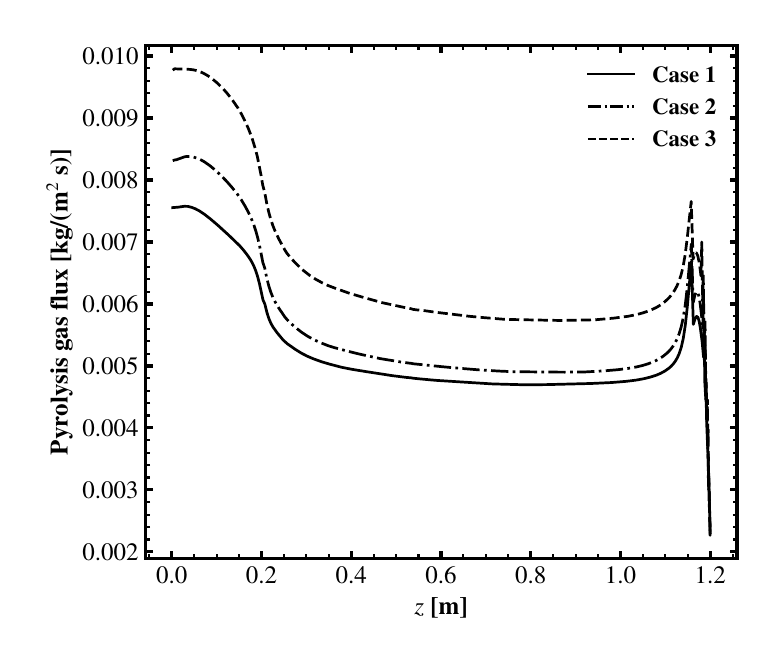}
    \caption{Pyrolysis gas flux}
  \end{subfigure}
  \caption{Wall distributions at $t = 111.1$~s for the equilibrium  model, Cases~1--3.}
  \label{fig:surf_eq}
\end{figure}

Figure~\ref{fig:surf_eq} compares Cases~1--3 of the equilibrium model at 111.1~s, which is the last analysis point and is obtained with the 35~km freestream conditions. The wall heat flux and the char mass flux differ little among the cases across the entire surface. The pyrolysis gas
flux, in contrast, is approximately 30\% higher in Case~3 than in Case~1 near the stagnation point and 20--25\% higher along the cone, because recession shortens the distance between the surface and the pyrolysis front. The increase is smaller in Case~2, approximately 10\% near the stagnation point and less than 5\% along the cone, because char recession is concentrated near the nose, whereas particle erosion produces a nearly uniform recession over the forebody (Fig.~\ref{fig:particle_surface}c). The wall temperature is correspondingly lower, because the receding surface exposes cooler material and steepens the temperature gradient near the surface. At the stagnation point, the wall temperature decreases from 1273~K in Case~1 to approximately 1250~K in Case~2 and 1218~K in Case~3, and Case~3 remains approximately 50~K below Case~1 along the cone. The effect of the surface recession treatment therefore appears in the blowing mass injection rather than in the surface heating.

\begin{figure}[htbp]
  \centering
  \begin{subfigure}[t]{0.49\textwidth}
    \centering
    \includegraphics[width=\linewidth]{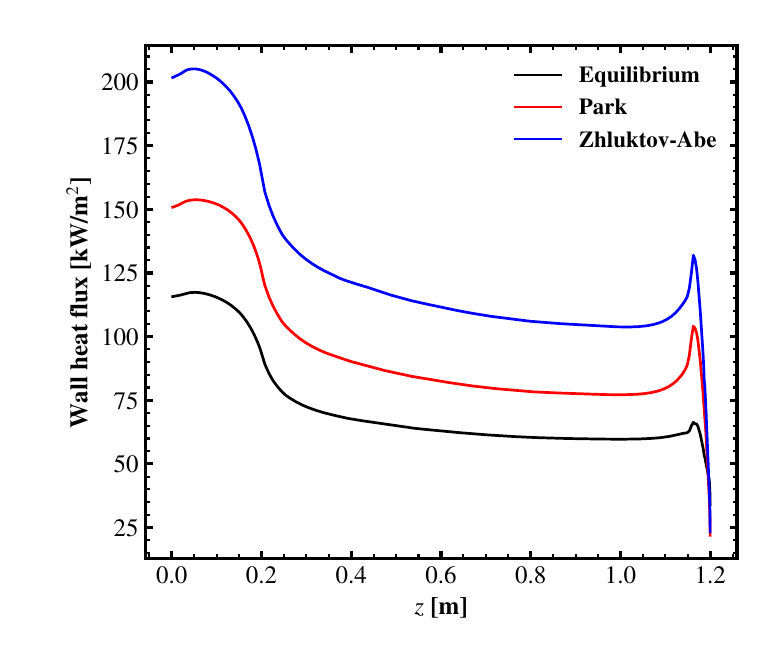}
    \caption{Wall heat flux}
  \end{subfigure}
  \hfill
  \begin{subfigure}[t]{0.49\textwidth}
    \centering
    \includegraphics[width=\linewidth]{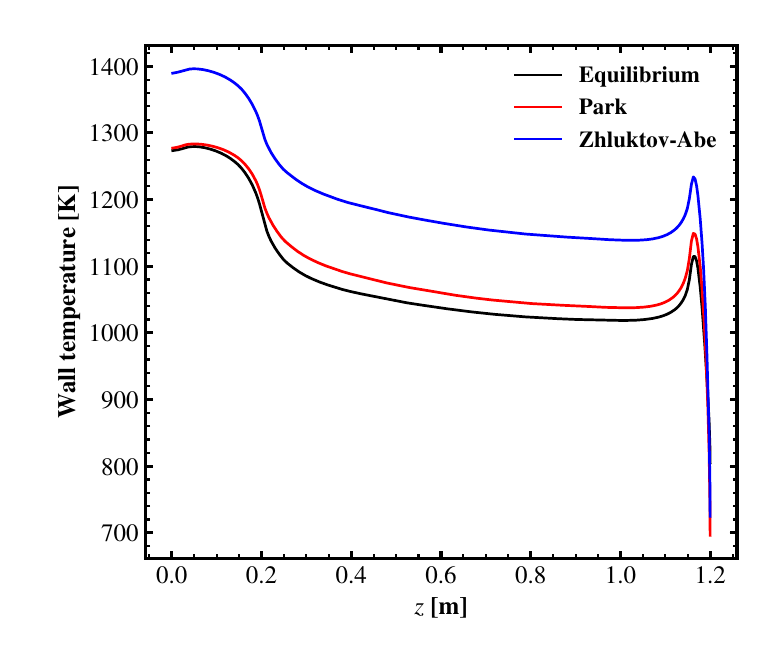}
    \caption{Wall temperature}
  \end{subfigure}

  \vspace{1ex}
  \begin{subfigure}[t]{0.49\textwidth}
    \centering
    \includegraphics[width=\linewidth]{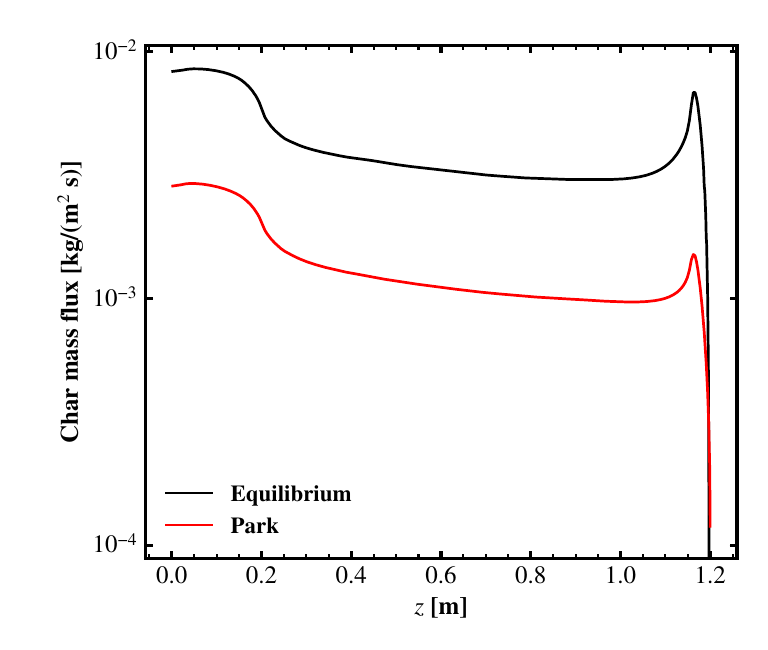}
    \caption{Char mass flux}
  \end{subfigure}
  \hfill
  \begin{subfigure}[t]{0.49\textwidth}
    \centering
    \includegraphics[width=\linewidth]{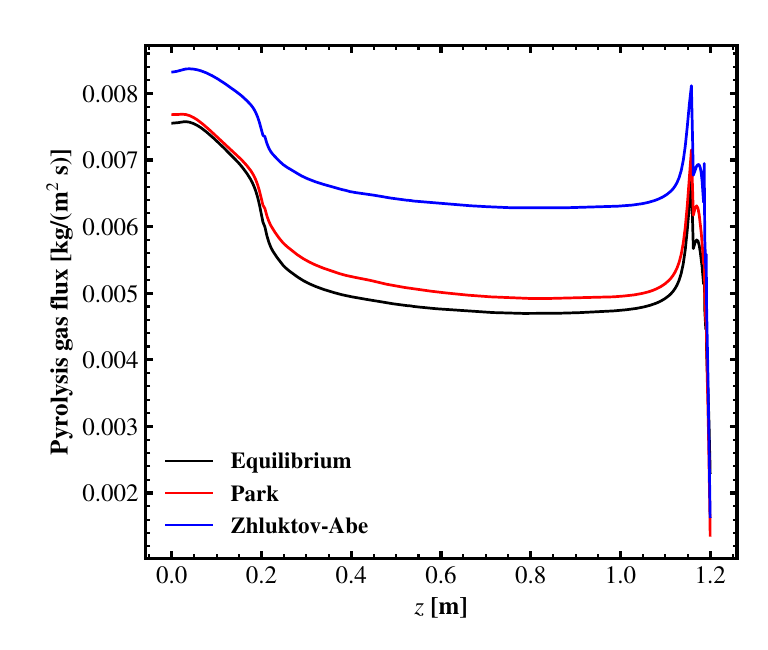}
    \caption{Pyrolysis gas flux}
  \end{subfigure}
  \caption{Wall distributions at $t = 111.1$~s for the three surface chemistry models, Case~1. The char mass flux of the ZA model is below the plotted range.}
  \label{fig:surf_mod}
\end{figure}

Figure~\ref{fig:surf_mod} compares the three surface chemistry models for Case~1 at the same time. Over most of the forebody, the equilibrium model predicts the lowest wall heat flux and temperature, followed by Park-5 and ZA. At the stagnation point, the wall heat flux is approximately 115, 152, and 205~kW/m$^2$ for the equilibrium, Park-5, and ZA models, respectively, and the same ordering holds along the cone. This ordering is the reverse of that of the char mass flux, which is approximately $8.3\times10^{-3}$ and $2.9\times10^{-3}$~kg/(m$^2$~s) at the stagnation point for the equilibrium and Park-5 models (Fig.~\ref{fig:surf_mod}c). The larger char injection increases the blowing at the wall and reduces the wall heat flux, consistent with the stagnation point comparison in Section~\ref{subsec:stag}. The wall temperatures of the equilibrium and Park-5 models are nearly equal at the stagnation point, approximately 1280~K, whereas the ZA model predicts approximately 1395~K. The pyrolysis gas flux shows a similar ordering, with the ZA model approximately 10\% higher at the stagnation point and 30\% higher along the cone than the equilibrium model. Although the material decomposition model is identical in all three cases, differences in surface heating and temperature history affect decomposition within the material. In contrast, the ZA model predicts a nearly zero char mass flux, resulting in negligible chemical ablation under the present conditions.

Overall, the results show distinct effects of surface chemistry, geometric recession, and particle interactions.  These effects are not independent: recession affects the in-depth decomposition, and particle erosion
enters the same chain by adding to that recession. Advancing the particle analysis and the material response with the flow therefore captures interactions that are absent when they are treated separately.

\FloatBarrier
\section{Conclusion}
  \label{sec:conclusion}

A coupled flow-material-particle framework was developed for Mars entry through a dusty atmosphere and applied to the Schiaparelli trajectory. The two-way flow-particle coupling previously developed by the authors was extended to a receding and pyrolyzing heat shield surface. The gas-surface interaction and the in-depth material response were solved separately on their own time scales, with the surface balances exchanged at every flow step and the in-depth response coupled implicitly over physical time windows. The particle phase was coupled loosely with explicit updates. Four consecutive trajectory segments between 50 and 30~km were computed using three surface chemistry models and three configurations: fixed geometry, char-induced recession, and combined char ablation and particle erosion. At the particle scale, the flow is in the transitional and free-molecular regimes, with particle Knudsen numbers of approximately 0.2 to 100, whereas the flow around the vehicle is a continuum. The Henderson drag correlation represents the rarefaction and compressibility effects in this regime. Because particle vaporization is negligible under the present conditions, mechanical erosion is governed by the impact velocity determined by drag. The present study also provides, to the authors' knowledge, the first quantitative comparison of the particle erosion predicted with rarefied and continuum drag correlations. Replacing the Henderson correlation with a continuum drag correlation reduces the stagnation point erosion rate by 31--35\%, because the smaller particles decelerate more strongly and the 1 and 2~$\mu$m particles do not reach the wall.

Surface chemistry influenced wall heat flux and char injection. Its effect on heat flux was larger than that of the geometric treatment. At 111.1~s, the stagnation point heat flux differed by approximately 30\% between the two finite-rate models and by more than 70\% between the Zhluktov--Abe and equilibrium models, compared with less than 1\% between the fixed and receding surfaces for the equilibrium model. The relative ordering of the equilibrium and Park-5 predictions changed along the trajectory, accompanying different trends in char injection. The Zhluktov--Abe model predicted the highest wall heat flux and temperature while producing negligible char ablation under the conditions examined. Including particle effects substantially increased material loss. The stagnation point recession at 111.1~s more than doubled for all three chemistry models. The particle erosion contribution remained close to 1.73~mm across the three chemistry models, whereas the char ablation contribution varied considerably. Surface recession had a comparatively small effect on gas-side heat flux but changed wall temperature and pyrolysis gas injection. The configurations with recession predicted lower wall temperatures and higher pyrolysis gas fluxes than the fixed geometry configuration. These differences reflect the interaction between the evolving geometry and the transient material response. The results support coupling the flow, surface balances, material response, and particle phase. The framework captures these interactions within a single calculation and evaluates their combined effects on wall heat flux, temperature, and cumulative recession along the flight trajectory.

\section*{CRediT authorship contribution statement}

\textbf{Hyeon Woo Nam:} Methodology, Software, Investigation, Formal analysis, Visualization, Writing -- original draft. \textbf{Sung Min Jo:} Conceptualization, Methodology, Supervision, Funding acquisition, Writing -- review \& editing.

\section*{Declaration of competing interest}

The authors declare that they have no known competing financial interests or personal relationships that could have appeared to influence the work reported in this paper.

\section*{Data availability}

The data that support the findings of this study are available from the corresponding author upon reasonable request. PATO, OpenFOAM, and preCICE are open-source software. HEGEL, PLATO, and ORACLE are research codes that are not publicly available.

\section*{Acknowledgements}

This work was partially supported by NASA under Award No.\ 80NSSC26M0171 and the MUREP Partnership Learning Annual Notification (MPLAN) Prize Challenge. This work was also partially supported by the AFOSR under Award No.\ FA9550-25-1-0119. The authors would like to thank Dr.\ A.~Munaf\`{o} and Dr.\ M.~Panesi (University of California, Irvine) for providing access to the PLATO and HEGEL software.

\appendix
  \section{Finite-rate surface chemistry models}
  \label{app:surface}
  This appendix describes the two finite-rate surface chemistry models applied to the surface, the Park-5 and Zhluktov--Abe (ZA) models.
  The surface notation is as follows: C(s) denotes solid carbon, (s) denotes a vacant active site, and O(s) and N(s) denote adsorbed oxygen and nitrogen atoms, respectively. In both models, $\dot{s}_i$ denotes the mass production rate of species $i$ per unit area, negative for species consumed at the wall, and the char mass flux is
\begin{equation}
    \dot{m}_c''=\sum_i \dot{s}_i.
\end{equation}

  \subsection{Park-5 model}

  The Park-5 model~\cite{johnston2014study} includes carbon oxidation by O and O$_2$, CN formation by atomic nitrogen, nitrogen recombination, and carbon sublimation as C$_3$.
  Table~\ref{tab:park5} summarizes the reactions and rate expressions.

  For the gas-surface reactions, the production rates are calculated from the incident species flux and the reaction probability $\gamma_j$. The net C$_3$ sublimation rate depends on the difference between the equilibrium vapor density and the local C$_3$ density at the wall.
  Here, $M_i$ is the molar mass. The coefficients are taken from Johnston~\cite{johnston2014study}, as also used by Le Maout et al.~\cite{le2025numerical}.

\begin{sidewaystable}
\centering
\caption{Park-5 surface reaction model~\cite{le2025numerical,johnston2014study}.}\label{tab:park5}
\scriptsize
\begin{tabular}{@{}llll@{}}
\toprule
Reaction $j$ & Type & Production rate & Parameters \\
\midrule
1.\ O + C(s) $\rightarrow$ CO & Eley--Rideal &
$\dot{s}_{\mathrm{CO}} = \rho_w Y_{w,\mathrm{O}} \gamma_1 \sqrt{\dfrac{R_u T_w}{2\pi M_{\mathrm{O}}}} \dfrac{M_{\mathrm{CO}}}{M_{\mathrm{O}}}$ &
$\gamma_1 = 0.63\, e^{-1160/T_w}$ \\[10pt]
2.\ O$_2$ + 2C(s) $\rightarrow$ 2CO & Eley--Rideal &
$\dot{s}_{\mathrm{CO}} = 2\rho_w Y_{w,\mathrm{O_2}} \gamma_2 \sqrt{\dfrac{R_u T_w}{2\pi M_{\mathrm{O_2}}}} \dfrac{M_{\mathrm{CO}}}{M_{\mathrm{O_2}}}$ &
$\gamma_2 = \dfrac{0.00143 + 0.01\, e^{-1450/T_w}}{1 + 0.0002\, e^{13000/T_w}}$ \\[10pt]
3.\ N + C(s) $\rightarrow$ CN & Eley--Rideal &
$\dot{s}_{\mathrm{CN}} = \rho_w Y_{w,\mathrm{N}} \gamma_3 \sqrt{\dfrac{R_u T_w}{2\pi M_{\mathrm{N}}}} \dfrac{M_{\mathrm{CN}}}{M_{\mathrm{N}}}$ &
$\gamma_3 = 0.001$ \\[10pt]
4.\ N + N(s) $\rightarrow$ N$_2$ + (s) & Eley--Rideal &
$\dot{s}_{\mathrm{N_2}} = \rho_w Y_{w,\mathrm{N}} \gamma_4 \sqrt{\dfrac{R_u T_w}{2\pi M_{\mathrm{N}}}}$ &
$\gamma_4 = 0.05$ \\[10pt]
5.\ 3C(s) $\rightarrow$ C$_3$ & Sublimation &
$\dot{s}_{\mathrm{C_3}} = \sqrt{\dfrac{R_u T_w}{2\pi M_{\mathrm{C_3}}}}\, \alpha_{\mathrm{C_3}} \left( A_{\mathrm{C_3}} T_w^{N_{\mathrm{C_3}}} e^{-E_{\mathrm{C_3}}/T_w} - \rho_w Y_{w,\mathrm{C_3}} \right)$ &
\begin{tabular}[t]{@{}l@{}}$\alpha_{\mathrm{C_3}} = 0.03$\\ $A_{\mathrm{C_3}} = 4.3197\times10^{22}$\\ $N_{\mathrm{C_3}} = -3.459$\\ $E_{\mathrm{C_3}} = 103{,}339$~K\end{tabular} \\
\bottomrule
\end{tabular}
\end{sidewaystable}

 \subsection{Zhluktov--Abe model}

  The Zhluktov--Abe (ZA) model~\cite{zhluktov1999viscous} represents the carbon surface with a fixed active-site density of $B=3.5\times10^{19}$~m$^{-2}$. Each site is either vacant or occupied by an adsorbed oxygen or nitrogen atom. The corresponding site fractions, $\Theta^0$, $\Theta_{\mathrm{O}}$, and $\Theta_{\mathrm{N}}$, respectively, satisfy
  \begin{equation}
      \Theta^0+\Theta_{\mathrm{O}}+\Theta_{\mathrm{N}}=1.
  \end{equation}
  They are evaluated from the steady balances of O(s) and N(s) at each wall state.

  The model contains twelve reversible reactions, listed in Table~\ref{tab:za}. Reverse rates are related to forward rates through the equilibrium constants $K_j$.
  These constants are constructed from the adsorption constants $K_1$ and $K_{11}$, the carbon vapor-pressure relation $K_8$, and gas-phase dissociation constants.
  The present calculations use the mobile-adsorption parameter set with high activation energies.
  The carbon vapor-pressure relation follows Blottner~\cite{blottner1970prediction}.

  The rate expressions in Table~\ref{tab:za} contain
\begin{equation}
    f_i=\frac{P_0}{\sqrt{2\pi m_i k T_w}},
\end{equation}
where $m_i$ is the mass of one gas particle of species $i$, $k$ and $h$ are the Boltzmann and Planck constants, and $P_0=1$~atm. In the rate expressions,  $p_w$ is the wall pressure in atm, $X_{w,i}$ is the wall mole fraction, and $\varepsilon_j$ is the factor in the $j$th forward or backward heterogeneous rate constant~\cite{zhluktov1999viscous}.

With the site fractions determined above, $\dot{s}_i$ follows from the reaction rates as in Ref.~\cite{zhluktov1999viscous}. In terms of these rates, the char mass flux can be obtained as follows: 
\begin{equation}
    \dot{m}_c'' = m_{\mathrm{C}}\left(r_5+r_6+r_7+r_8+2r_9+3r_{10}\right).
\end{equation}

\begin{table}[h]
\centering
\caption{Zhluktov--Abe surface reaction model~\cite{zhluktov1999viscous}.}\label{tab:za}
\renewcommand{\arraystretch}{1.35}
\scriptsize
\setlength{\tabcolsep}{2pt}
\begin{tabular}{@{}llcc@{}}
\toprule
Reaction $j$ & Net rate, $r_j$ & $\varepsilon_j$ & $T_{aj}$ (K) \\
\midrule
1.\ O + (s) $\rightleftharpoons$ O(s) & $\varepsilon_1 f_{\mathrm{O}}\left(p_w X_{w,\mathrm{O}}\Theta^0 - \Theta_{\mathrm{O}}/K_1\right)$ & $1$ & -- \\
2.\ O$_2$ + 2(s) $\rightleftharpoons$ 2O(s) & $\varepsilon_2 B\frac{kT_w}{h} e^{-T_{a2}/T_w}\left[K_2 p_w X_{w,\mathrm{O_2}}(\Theta^0)^2 - \Theta_{\mathrm{O}}^2\right]$ & $8\times10^{-4}$ & $30{,}800$ \\
3.\ O$_2$ + (s) $\rightleftharpoons$ O(s) + O & $\varepsilon_3 f_{\mathrm{O_2}} e^{-T_{a3}/T_w}\left(p_w X_{w,\mathrm{O_2}}\Theta^0 - p_w X_{w,\mathrm{O}}\Theta_{\mathrm{O}}/K_3\right)$ & $1$ & $14{,}200$ \\
4.\ CO$_2$ + (s) $\rightleftharpoons$ O(s) + CO & $\varepsilon_4 f_{\mathrm{CO_2}}\left(p_w X_{w,\mathrm{CO_2}}\Theta^0 - p_w X_{w,\mathrm{CO}}\Theta_{\mathrm{O}}/K_4\right)$ & $0.9$ & -- \\
5.\ O(s) + C(s) $\rightleftharpoons$ CO + (s) & $\varepsilon_5 B\frac{kT_w}{h} e^{-T_{a5}/T_w}\left(\Theta_{\mathrm{O}} - p_w X_{w,\mathrm{CO}}\Theta^0/K_5\right)$ & $0.1$ & $40{,}000$ \\
6.\ O + O(s) + C(s) $\rightleftharpoons$ CO$_2$ + (s) & $\varepsilon_6 f_{\mathrm{O}} e^{-T_{a6}/T_w}\left(p_w X_{w,\mathrm{O}}\Theta_{\mathrm{O}} - p_w X_{w,\mathrm{CO_2}}\Theta^0/K_6\right)$ & $0.8$ & $2000$ \\
7.\ 2O(s) + C(s) $\rightleftharpoons$ CO$_2$ + 2(s) & $\varepsilon_7 B\frac{kT_w}{h} e^{-T_{a7}/T_w}\left[\Theta_{\mathrm{O}}^2 - p_w X_{w,\mathrm{CO_2}}(\Theta^0)^2/K_7\right]$ & $1$ & $40{,}000$ \\
8.\ C(s) $\rightleftharpoons$ C & $\varepsilon_8 f_{\mathrm{C}}\,\Theta^0\left(K_8 - p_w X_{w,\mathrm{C}}\right)$ & $0.24$ & -- \\
9.\ 2C(s) $\rightleftharpoons$ C$_2$ & $\varepsilon_9 f_{\mathrm{C_2}}(\Theta^0)^2\left(K_9 - p_w X_{w,\mathrm{C_2}}\right)$ & $0.5$ & -- \\
10.\ 3C(s) $\rightleftharpoons$ C$_3$ & $\varepsilon_{10} f_{\mathrm{C_3}}(\Theta^0)^3\left(K_{10} - p_w X_{w,\mathrm{C_3}}\right)$ & $0.023$ & -- \\
11.\ N + (s) $\rightleftharpoons$ N(s) & $\varepsilon_{11} f_{\mathrm{N}}\left(p_w X_{w,\mathrm{N}}\Theta^0 - \Theta_{\mathrm{N}}/K_{11}\right)$ & $1$ & -- \\
12.\ N + N(s) $\rightleftharpoons$ N$_2$ + (s) & $\varepsilon_{12} f_{\mathrm{N_2}} e^{-T_{a12}/T_w}\left(K_{12} p_w X_{w,\mathrm{N}}\Theta_{\mathrm{N}} - p_w X_{w,\mathrm{N_2}}\Theta^0\right)$ & $1$ & $76{,}600$ \\
\bottomrule
\end{tabular}
\end{table}

\bibliographystyle{elsarticle-num}
\bibliography{references}

\end{document}